\documentclass[aps,twocolumn,superscriptaddress,amsmath,longbibliography,nofootinbib]{revtex4-1}
\usepackage{graphicx,color,hyperref,xcolor}
\usepackage{soul}
\usepackage{amsfonts}
\usepackage{amssymb}
\usepackage{amsmath}
\usepackage{amsthm}
\usepackage{bm}
\usepackage{braket}
\usepackage{multirow}
\usepackage{array}

\makeatletter
\newcommand*{\rom}[1]{\expandafter\@slowromancap\romannumeral #1@}
\makeatother

\begin{document}

\title{Microwave Microscope Studies of Trapped Vortex Dynamics in Superconductors}

\author{Chung-Yang Wang}
\affiliation{Quantum Materials Center, Department of Physics, University of Maryland, College Park, Maryland 20742, USA}

\author{Steven M. Anlage}
\affiliation{Quantum Materials Center, Department of Physics, University of Maryland, College Park, Maryland 20742, USA}

\begin{abstract}

Trapped vortices in superconductors introduce residual resistance in superconducting radio-frequency (SRF) cavities and disrupt the operation of superconducting quantum and digital electronic circuits. Understanding the detailed dynamics of trapped vortices under oscillating magnetic fields is essential for advancing these technologies. We have developed a near-field magnetic microwave microscope to study the dynamics of a limited number of trapped vortices under the probe when stimulated by a localized rf magnetic field. By measuring the local second-harmonic response ($P_\mathrm{2f}$) at sub-femto-Watt levels, we isolate signals exclusively arising from trapped vortices, excluding contributions from surface defects and Meissner screening currents. Toy models of Niobium superconductor hosting vortex pinning sites are introduced and studied with Time-Dependent Ginzburg-Landau (TDGL) simulations of probe/sample interaction to better understand the measured second-harmonic response. The simulation results demonstrate that the second-harmonic response of trapped vortex motion under a localized rf magnetic field shares key features with the experimental data. This measurement technique provides access to vortex dynamics at the micron scale, such as depinning events and spatially-resolved pinning properties, as demonstrated in measurements on a Niobium film with an antidot flux pinning array.

\end{abstract}

\maketitle



\section{Introduction}
\label{sec:Introduction}

Stray magnetic vortices that become pinned in superconductors contribute to enhanced losses in superconducting radio-frequency (SRF) cavities and corrupt the operation of superconducting electronics. In SRF cavities, trapped vortices introduce residual resistance, limiting the achievable quality factor \cite{Benvenuti1999Study,Ciovati2008Evidence,Gurevich2008Dynamics,Aull2012Trapped,Gurevich2013Effect,Romanenko2014Ultra,Romanenko2014Dependence,Gonnella2016Impact,Posen2016Efficient,Dhakal2017Effect,Padamsee2017Years,Checchin2017Electron,Liarte2018Vortex,carlson2021analysis,Khanal2024Role}. For superconducting quantum circuits, trapped vortices add loss \cite{Song2009Reducing,Song2009Microwave,Eley2021Challenges}, which serve to limit the coherence time of superconducting qubits operating at microwave frequencies. On the other hand, trapped vortices at current-nodes of a superconducting resonator can also be used to capture non-equilibrium quasiparticles and relax them \cite{Nsanzineza2014Trapping}. For superconducting digital circuits, pinned vortices not only degrade performance but, in extreme cases, can compromise circuit functionality entirely \cite{Bermon1983Moat,Jeffery1995Magnetic,Suzuki2006Investigation,Polyakov2007Flux,Narayana2009Evaluation,Fujiwara2009Research,Semenov2016How,Jackman2017Flux,Uday2022Superconducting,Schindler2024TheEffect}. The effects of trapped vortices in superconducting digital circuits are usually thought of in terms of DC bias currents that the vortices add to the circuits. However, in this work we investigate an effect that has largely been overlooked in the context of superconducting digital circuits, namely the microwave frequency response of trapped vortices in superconductors.

The behavior of trapped vortices under stimulation by an oscillating magnetic field \cite{Pathirana2021Effect} is often studied macroscopically, focusing on the average properties of many trapped vortices, such as loss, critical current density, and pinning energy \cite{Golosovsky1994Vortex,Moshchalkov1998Pinning,Raedts2004FluxPinning,Silhanek2005Enhanced,Cuadra2015rfcoil,dobrovolskiy2020moving,Nakamura2020Nonreciprocal,Alimenti2020Microwave}. These measurements are generally interpreted in terms of collective single-coordinate models (e.g., Gittleman-Rosenblum type models) \cite{Gittleman1966RadioFrequency,Coffey1991Unified,Pambianchi1993DC,Wu1995Frequency,Golosovsky1996High,Enrico2017Vortices}, which subsumes the response of individual vortices to the rf currents. While these models provide insights into the average properties of all trapped vortices in a sample, direct high-frequency microscopic measurements at the level of individual trapped vortices remain challenging. Furthermore, beyond measuring loss, it is essential to understand the detailed dynamics of trapped vortices under oscillating magnetic fields. To achieve this, we need to study small numbers of vortices pinned in superconductors to understand their full dynamics and pinning properties.

Here we introduce a near-field magnetic microwave microscope approach \cite{lee2000magnetic,lee2003Study,lee2003spatially,lee2005doping,lee2005microwave,mircea2009phase,tai2011nonlinear,tai2012nanoscale,tai2014near,tai2014modeling,tai2015nanoscale,oripov2019high,wang2024microscopic,Wang2024Near} to study the dynamics of a limited number of trapped vortices under the probe (at micron length scales) by stimulating them with a localized and intense rf magnetic field and analyzing the second-harmonic response ($P_\mathrm{2f}$) generated by their resulting motion. 

Various microscopic techniques have been employed to study individual vortices in superconductors, including magnetic force microscopy (MFM) \cite{Dremov2019Local}, scanning superconducting quantum interference device (SQUID) \cite{Embon2015Probing,Bishop2023Vortex}, single-vortex confinement structures \cite{Marek2024Quantum}, scanning Hall probe microscopy \cite{Zhang2019Direct}, and scanning tunneling microscopy (STM) \cite{Chen2024Revealing}. These methods have provided invaluable insights into vortex behavior. However, they do not directly probe the microwave response of vortices. In contrast, our approach employs a near-field magnetic microwave microscope to investigate the high-frequency response of trapped vortices, enabling the study of their dynamics under localized microwave excitation.

The response of a superconductor to an oscillating magnetic field can be classified by symmetry into odd harmonic responses (linear response, third-harmonic response, etc.) and even harmonic responses (second-harmonic response, fourth-harmonic response, etc.), which probe fundamentally different aspects of superconducting dynamics \cite{Jeffries1988Symmetry,Ji1989Critical,Muller1989Nonlinear,Jeffries1989Nonlinear,Ishida1990Fundamental,Yamamoto1992Harmonic,Shatz1993Universal,Portis1993Electrodynamics,Samoilova1995Nonlinear,Oates2001Nonlinear}. Odd harmonic responses \cite{lee2000magnetic,lee2003spatially,lee2005doping,lee2005microwave,Velichko2005Nonlinear,mircea2009phase,tai2011nonlinear,tai2012nanoscale,tai2014near,tai2014modeling,tai2015nanoscale} capture contributions from a variety of mechanisms, including Meissner screening currents \cite{Oates2004Observation,Zhuravel2013Imaging}, nonlinearities from current-dependent superfluid density variation, and rf vortex nucleation \cite{oripov2019high,wang2024microscopic}. In contrast, even harmonic responses \cite{lee2003spatially,lee2005doping,lee2005microwave,Nakamura2020Nonreciprocal} are known to vanish in the absence of time reversal invariance breaking \cite{Jeffries1988Symmetry,Muller1989Nonlinear,Jeffries1989Nonlinear,Ishida1990Fundamental}, such as when no DC magnetic field is applied and no vortices are trapped in the superconductor. Notably, in the absence of an external offset (field or current), the second-harmonic response collects signals from trapped vortices exclusively, effectively filtering out contributions from other mechanisms. This exclusivity makes $P_\mathrm{2f}$ a powerful tool for studying trapped vortices \cite{Nakamura2020Nonreciprocal}.

In previous research, we demonstrated that our microwave microscope can study surface defects that nucleate rf vortices by measuring the \textit{third}-harmonic response $P_\mathrm{3f}$ \cite{wang2024microscopic}. These rf vortices are nucleated and destroyed twice in each rf cycle \cite{Gurevich2008Dynamics,oripov2020time}. In this work, we show that the same microwave microscope can be used to instead study trapped DC vortices by measuring the \textit{second}-harmonic response, $P_\mathrm{2f}$. 

However, the second-harmonic response has some subtleties associated with its interpretation. We show that the magnitude of the second-harmonic response is \textit{not} directly proportional to the density of trapped vortices, or the surface impedance of the sample. Note that it is not our goal to measure the critical current density or the vortex depinning frequency (which is an effective macroscopic property of many vortices) associated with trapped DC vortices. Instead, $P_\mathrm{2f}$ is sensitive to a limited number of vortices and their detailed configuration and dynamics beneath the probe of our microwave microscope, and hence can be used to detect \textit{changes in} trapped vortex configuration. 

In this work, we experimentally study a Nb film with an antidot vortex pinning array. Our results suggest that this measurement technique is useful for detecting depinning events of trapped vortices at the micron scale and studying spatially-resolved pinning properties. 

The structure of this paper is as follows: Sec. \ref{sec:Setup} describes the experimental setup. In Sec. \ref{sec:TDGL}, we discuss the mechanism underlying the second-harmonic response observed in our measurements and use Time-Dependent Ginzburg-Landau (TDGL) simulations to highlight three key features of the second-harmonic response associated with trapped vortex dynamics. Sec. \ref{sec:Results} presents the experimental results obtained from the Nb film. Finally, in Sec. \ref{sec:Conclusion}, we summarize our findings and discuss the capabilities of this measurement technique.


\section{Experimental setup}
\label{sec:Setup}

Near-field microwave microscopy of materials properties \cite{TabibAzar1993Nondestructive,Vlahacos1996Near,Steinhauer1997Surface,Steinhauer1998Quantitative,Vlahacos1998Quantitative,TabibAzar1999Nondestructive,Anlage1999Superconducting,Anlage2001Near,Rosner2002High,Atif2003Anovel,Tselev2003Near,Imtiaz2005Near,Imtiaz2006Effect,Anlage2007Principles,Imtiaz2007Nanometer,Tselev2007Broadband,barber2022microwave} has also proven to be very helpful in the study of superconducting microwave devices \cite{Anlage1997Scanning,Thanawalla1998Microwave,Hu1999Imaging,Anlage2021Microwave}. The setup of our near-field magnetic microwave microscope is identical to that described in Ref. \cite{Wang2024Near,wang2024microscopic}. A brief overview of the microscope is provided in this section. 

The core component of our microwave microscope is a magnetic writer head (provided by Seagate Technology), commonly used in conventional hard-disk drives. This magnetic writer head can generate a localized rf magnetic field. In our setup, a Seagate magnetic writer head is mounted on a cryogenic XYZ positioner and operated in a scanning probe microscope configuration.

The microwave source (HP 83620B) signal $P_\mathrm{rf}\mathrm{sin}^{2}(\omega t)$ is sent to the probe (magnetic writer head) via its built-in transmission line. The probe then produces a localized rf magnetic field $B_\mathrm{rf}\mathrm{sin}(\omega t)$ acting on the sample surface. Trapped vortices in the sample then wiggle and generate a response magnetic field that is coupled back to the same probe, creating a propagating signal whose second-harmonic component $P_\mathrm{2f}\mathrm{sin}^{2}(2 \omega t)$ is then extracted and measured by a spectrum analyzer (Aaronia RSA250X) at room temperature. In practice, a complicated multi-harmonic signal comes back to room temperature, but we filter out just the second-harmonic response.

With magnetic shielding in place, the residual DC magnetic field near the sample at low temperatures is found to be approximately 2.1 $\mu$T, as measured with an in-situ cryogenic 3-axis magnetometer (Bartington Cryomag-100).

The spatial resolution of the microscope ranges from sub-micron to micron scale, depending on the probe-sample separation, and the signal being analyzed. The spatial resolution for measurements of second-harmonic response is estimated to be approximately $1.1 \, \mu$m for this specific setup (see Appendix \ref{sec:SpatialResolution}).


\section{Numerical simulations}
\label{sec:TDGL}

The dynamics of trapped vortices under stimulation by a localized rf magnetic field is a complex topic. A comprehensive numerical investigation is beyond the scope of this work. Instead, this section aims to illustrate key features of $P_\mathrm{2f}$ associated with trapped vortex dynamics using numerical simulations of simplified toy models (Niobium superconductor hosting vortex pinning sites). These insights will later be used to analyze the experimental results (Sec. \ref{sec:Results}). 

In the following, we first introduce our simulation framework (Sec. \ref{sec:TDGLIntroduction}), then discuss the mechanism underlying the second-harmonic response observed in our measurements (Sec. \ref{sec:P2f}), and finally present three core features using various toy models (Sec. \ref{sec:Feature1}, Sec. \ref{sec:Feature2}, and Sec. \ref{sec:Feature3}).


\subsection{Introduction to numerical simulations}
\label{sec:TDGLIntroduction}

\begin{figure}
\includegraphics[width=0.4\textwidth]{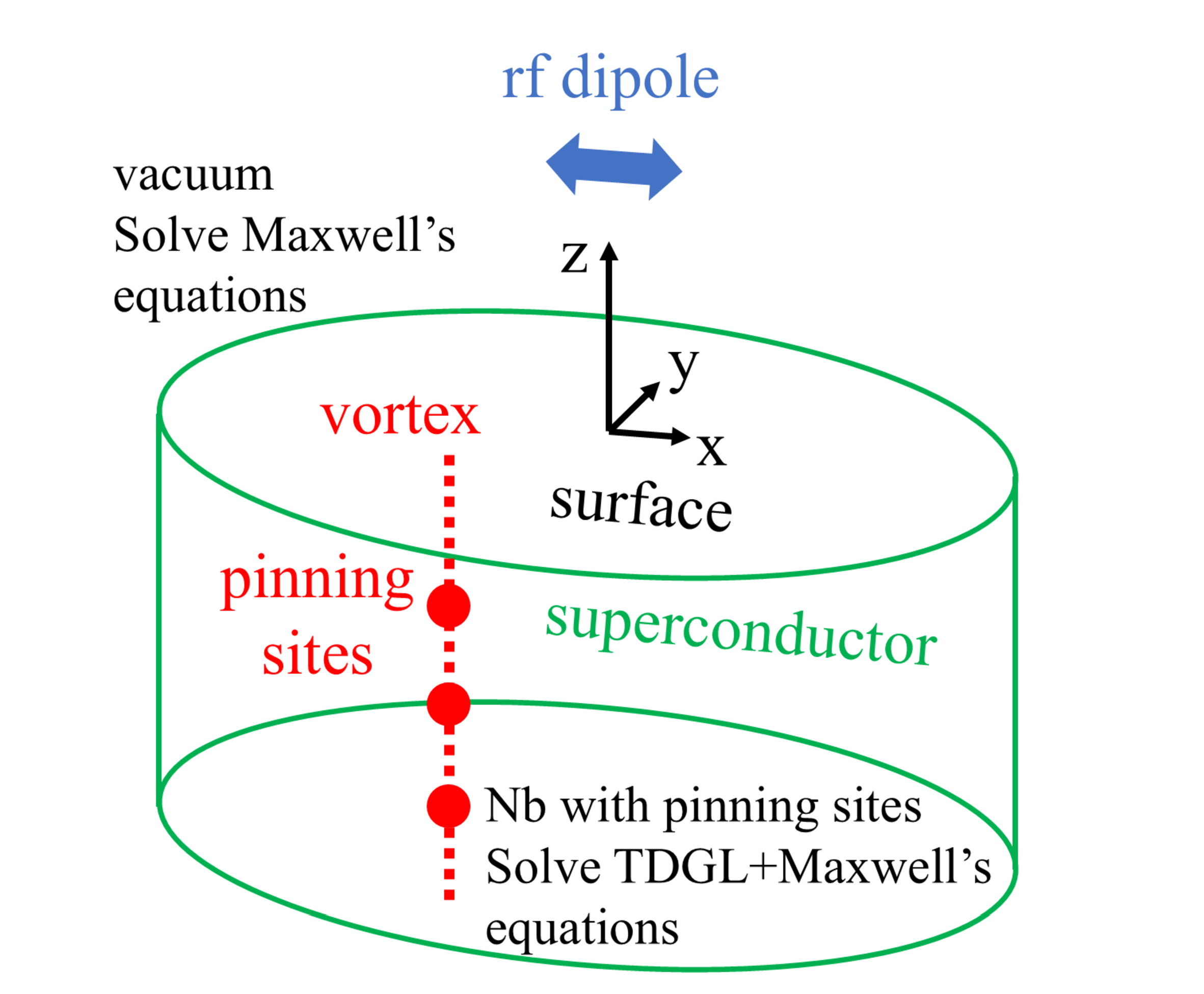}
\caption{\label{fig:TDGLSettingSchematics}A schematic representation of the TDGL simulations and toy models. The blue double arrow represents the rf dipole, the red dots represent the pinning sites, and the red dashed line represents the trapped vortex.}
\end{figure}

The Time-Dependent Ginzburg-Landau (TDGL) model is a widely used framework for studying vortex dynamics in superconductors \cite{kato1999charging,Deang2001Stochastic,Jennifer2002Modeling,hernandez2008dissipation,lara2015microwave,dobrovolskiy2020moving,pack2020vortex,carlson2021analysis,wang2022effects,Bogush2024Microwave}. In previous work, we employed TDGL simulations to investigate rf vortex nucleation and $P_\mathrm{3f}$ \cite{oripov2019high,oripov2020time,wang2024microscopic}. In this work, we use the same simulation framework to study trapped vortex dynamics and $P_\mathrm{2f}$. In the following, we provide a brief overview of the key settings used in the TDGL simulations, as the detailed descriptions are provided in Ref. \cite{wang2024microscopic}. The equations, boundary conditions, material parameters, and the setting of toy models of the TDGL simulations can be found in Appendix \ref{sec:TDGLsetting}.

Figure \ref{fig:TDGLSettingSchematics} provides a schematic representation of the TDGL simulations and toy models. In our measurements, the applied field generated by the microwave microscope probe is a localized rf magnetic field with a configuration similar to that produced by a point dipole aligned parallel to the sample surface. In the simulations, the probe is approximated as a pointlike magnetic dipole located at $(x_\mathrm{dp},y_\mathrm{dp},h_\mathrm{dp})=(0,0,400 \, \mathrm{nm})$, oriented in the x-direction, and having a sinusoidal time-dependent magnetic moment $(M_\mathrm{dp}\mathrm{sin}(\omega t),0,0)$ with a frequency of $\omega/{2\pi}$=1.7 GHz. To quantify the strength of the rf magnetic field, the peak rf magnetic field amplitude experienced by the superconductor is specified and is denoted as $B_\mathrm{pk}$. 

Note that the TDGL simulations are fully three-dimensional, hence quite demanding in terms of memory usage and processing time. The model is inhomogeneous in the sense that it consists of both a vacuum region, which hosts the rf dipole acting as a source of high frequency magnetic field, and a superconducting region which hosts the DC vortex (see Fig. \ref{fig:TDGLSettingSchematics}). Maxwell's equations are solved in the vacuum region, while the coupled TDGL equations, along with Maxwell's equations are solved inside the superconductor, and appropriate boundary conditions are enforced at the surfaces. 

The simulation domain is of finite size, and no periodic boundary conditions are imposed. Since the primary vortex dynamics occur near the origin, the influence of finite-size effects is mitigated by selecting a sufficiently large simulation domain and applying appropriate boundary conditions. Further details are provided in Appendix \ref{sec:TDGLsetting}.

The primary addition to the setup described in Ref. \cite{wang2024microscopic} is the inclusion of single-flux-quantum trapped magnetic vortices. To generate trapped vortices, spherical pinning sites composed of a low-$T_\mathrm{c}$ impurity phase are introduced into the Nb domain, and a uniform DC magnetic field is applied along the z-direction. DC vortices aligned in the z-direction then nucleate at the boundary of the superconducting domain and propagate inward, spreading throughout the superconducting region. While most of these vortices remain unpinned and free to move, some become pinned at the pinning sites. Once these vortices are pinned, the external DC magnetic field is subsequently \textit{removed}. The unpinned vortices repel each other and eventually exit the simulation domain, leaving only the pinned vortices localized at the pinning sites.  

After the unpinned vortices have left the simulation domain, the rf magnetic field is applied for five rf cycles, and $P_\mathrm{2f}$ at the location of the dipole is calculated based on the vortex dynamics observed during the fifth rf cycle.


\subsection{Origin of second-harmonic response}
\label{sec:P2f}

Consider a system driven by a sinusoidal driving force with angular frequency $\omega$. Let the response function be denoted as $f(t)$. The even components of the Fourier transform of $f(t)$ vanish if $f(t) = -f(t + \pi/\omega)$, meaning the response function exhibits perfect symmetry between the two half-cycles. In other words, even harmonic responses quantify the asymmetric component between the two half-cycles of $f(t)$. As the leading-order term of even harmonic responses, the second-harmonic response is frequently analyzed in such studies. 

In the context of this work, the system consists of trapped vortices in superconductors, and the driving force is an oscillating magnetic field created by a scanned dipole over the surface of the superconductor. When trapped vortices are stimulated by an oscillating magnetic field, they wiggle in response, and the asymmetric component of this motion generates even harmonic responses.

The asymmetry in time of trapped vortex dynamics can arise from various mechanisms. One such mechanism involves an asymmetric potential experienced by vortices, which may originate intrinsically in noncentrosymmetric superconductors \cite{Ryohei2017Nonreciprocal,Itahashi2020Quantum}, be introduced through artificially engineered asymmetric potentials \cite{Lee1999Reducing,Souza2006Vortex,Plourde2009Nanostructured,Jin2010High}, or be created by the combination of a symmetric pinning potential and a DC current acting as a bias \cite{Souza2006Vortex,Nakamura2020Nonreciprocal}. In our case, the asymmetry originates from the localized and inhomogeneous nature of the rf magnetic field generated by the microwave microscope. This rf magnetic field induces a screening current, which in turn generates a Lorentz force acting on trapped vortices \cite{Gittleman1966RadioFrequency,Coffey1991Unified,Golosovsky1996High,Enrico2017Vortices}. Due to the strong spatial gradient of the rf magnetic field, both the screening current and the resulting Lorentz force exhibit significant spatial gradients. During an rf cycle, a trapped vortex experiences a stronger Lorentz force when moving toward regions of higher magnetic field strength and a weaker force when moving away, resulting in asymmetric vortex dynamics between the two half-cycles.


\begin{figure}
\includegraphics[width=0.48\textwidth]{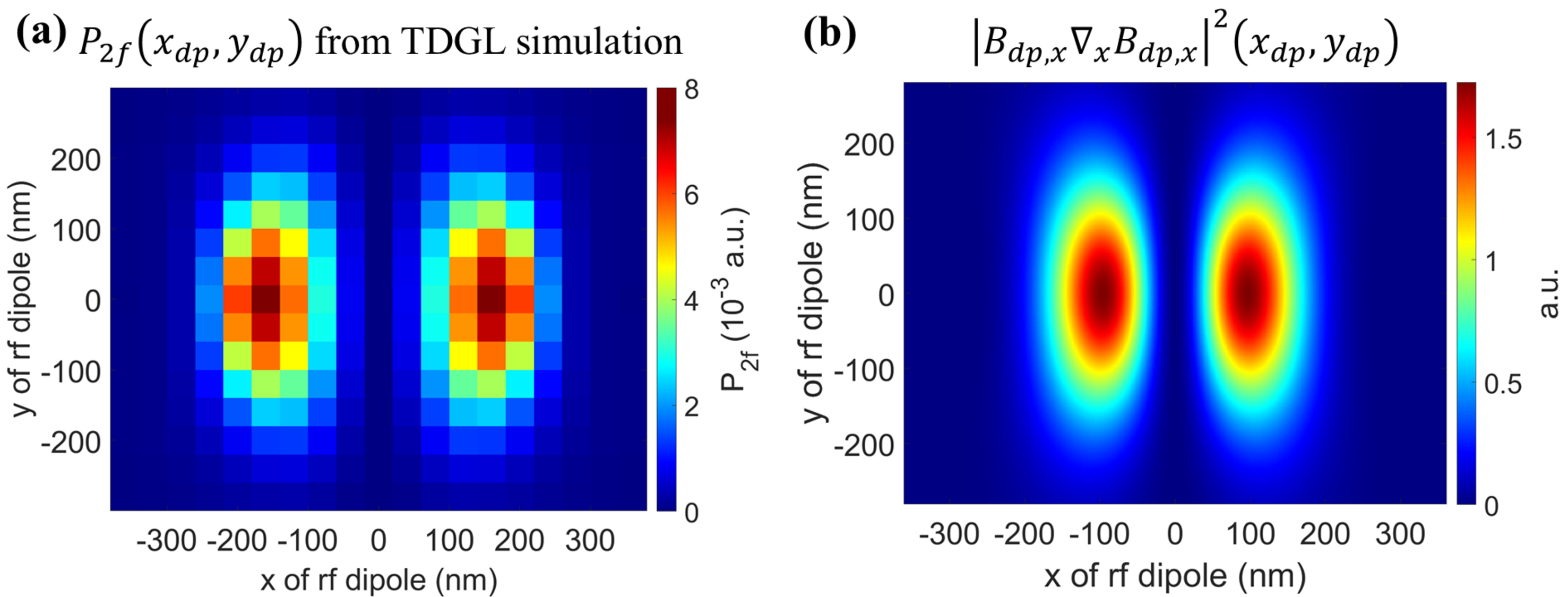}
\caption{\label{fig:DipoleFieldandP2f}Asymmetry in time of the dynamics of a trapped vortex (at $(x,y)=(0,0)$) as a function of dipole location $(x_\mathrm{dp}, y_\mathrm{dp})$, for $h_\mathrm{dp}=400 \, \mathrm{nm}$. (a) TDGL simulation results of $P_\mathrm{2f}$ as a function of dipole location $(x_\mathrm{dp}, y_\mathrm{dp})$ for toy model 1 (a vortex pinned at $(x,y) = (0,0)$), at $T=7.44$ K and $B_\mathrm{pk}=28.3$ mT. (b) The plot of ${\left| B_\mathrm{dp,x} \nabla_{x} B_\mathrm{dp,x} \right|}^2$ at $(x,y)=(0,0)$ as a function of dipole location $(x_\mathrm{dp}, y_\mathrm{dp})$, where $B_\mathrm{dp,x}$ is the x-component of the magnetic field at the surface of a superconductor created by a magnetic dipole.}
\end{figure}

It is informative to ask: For a vortex pinned at $(x,y)=(0,0)$, how does its dynamics vary as a function of the dipole location $(x_\mathrm{dp}, y_\mathrm{dp})$ with fixed height $h_\mathrm{dp}$? Figure \ref{fig:DipoleFieldandP2f}(a) shows the TDGL simulation results for $P_\mathrm{2f}$ (recovered at the location of the dipole) as a function of $(x_\mathrm{dp}, y_\mathrm{dp})$, for a vortex pinned at $(x,y) = (0,0)$ (toy model 1). It essentially represents the second-harmonic point-spread function of a single trapped DC vortex, as imaged by our near-field magnetic microwave microscope. The pattern of $P_\mathrm{2f}(x,y)$ shown in Fig. \ref{fig:DipoleFieldandP2f}(a) can be understood as follows.

The Lorentz force $F_\mathrm{Lorentz}$ acting on a trapped vortex is proportional to the local current density $J$, which, to first-order approximation, is proportional to the magnitude of the magnetic field $B_\mathrm{total}$. At the superconductor surface ($z=0$), $B_\mathrm{total} = 2 B_\mathrm{external}$ by considering the image method and the boundary condition that the perpendicular component of the magnetic field must vanish at $z=0$. In other words, at $z=0$, we have
\begin{equation} \label{eq:LorentzForce}
F_\mathrm{Lorentz} \propto J \propto B_\mathrm{total} = 2 B_\mathrm{external}.
\end{equation}
In our case, $B_\mathrm{external}$ is the external magnetic field at $z=0$ created by the rf dipole.

Since the rf dipole is oriented in the x-direction, the induced current is primarily in the y-direction, creating a Lorentz force on the vortex in the x-direction, and the motion of the trapped vortex is predominantly along the x-axis. Roughly speaking, the asymmetry in the vortex motion is proportional to both the strength of the Lorentz force exerted on the vortex by $J_\mathrm{y}$ and its spatial gradient at $z=0$. Consequently, an estimate of $P_\mathrm{2f}$ at the surface of the superconductor is given by
\begin{equation} \label{eq:DipoleField}
P_\mathrm{2f}(x_\mathrm{dp},y_\mathrm{dp}) \propto {\left| J_\mathrm{y} \nabla_{x} J_\mathrm{y} \right|}^2 \propto {\left| B_\mathrm{dp,x} \nabla_{x} B_\mathrm{dp,x} \right|}^2,
\end{equation}
where $B_\mathrm{dp,x}$ is the x-component of the dipole magnetic field ($B_\mathrm{external}$ in this case) at $z=0$. The analytic expressions for $B_\mathrm{dp,x}$ and $B_\mathrm{dp,x} \nabla_{x} B_\mathrm{dp,x}$ due to a magnetic dipole above the surface of a superconductor are provided in Appendix \ref{sec:DipoleBField}.

Figure \ref{fig:DipoleFieldandP2f}(b) presents ${\left| B_\mathrm{dp,x} \nabla_{x} B_\mathrm{dp,x} \right|}^2$ experienced at $(x,y)=(0,0)$ as a function of scanned dipole location $(x_\mathrm{dp}, y_\mathrm{dp})$. The strong similarity between Fig. \ref{fig:DipoleFieldandP2f}(a) and Fig. \ref{fig:DipoleFieldandP2f}(b) supports the qualitative validity of Eq. \ref{eq:DipoleField}. A key feature shared by both figures is that the asymmetry vanishes along the entire y-axis ($x_\mathrm{dp} = 0$). In Fig. \ref{fig:DipoleFieldandP2f}(b), this occurs because $\nabla_{x} B_\mathrm{dp,x} = 0$. In Fig. \ref{fig:DipoleFieldandP2f}(a), the system exhibits symmetry between the $-x$ and $+x$ directions, leading to symmetric vortex dynamics between the two half-cycles, which results in $P_\mathrm{2f} = 0$. Note that $P_\mathrm{2f} = 0$ when the dipole is positioned directly above the trapped vortex ($(x_\mathrm{dp}, y_\mathrm{dp}) = (0,0)$) even though there is a significant Lorentz force acting on the vortex. Figure \ref{fig:DipoleFieldandP2f}(a) clearly demonstrates that $P_\mathrm{2f}$ reflects the asymmetry of the vortex dynamics rather than simply indicating the presence of a trapped vortex right beneath the dipole. In the case of multiple trapped vortices, $P_\mathrm{2f}$ remains a measure of the asymmetry in their collective dynamics rather than being proportional to their number.

It is worth pointing out that the pinning potential in toy model 1 enjoys rotational symmetry about the z-axis. The asymmetry in the wiggling motion of the trapped vortex (except along the y-axis) originates from the strong gradient of the localized rf magnetic field itself at the vortex location rather than from any intrinsic asymmetry in the pinning potential. Consequently, this technique, measuring $P_\mathrm{2f}$ with our microwave microscope, remains effective even for trapped vortices in symmetric pinning potentials.

For the rest of the simulations (toy models 2 to 6), the dipole is fixed at $(x_\mathrm{dp},y_\mathrm{dp},h_\mathrm{dp})=(0,0,400 \, \mathrm{nm})$.


\subsection{Trapped vortices and $P_\mathrm{2f}$}
\label{sec:Feature1}

\begin{figure}
\includegraphics[width=0.45\textwidth]{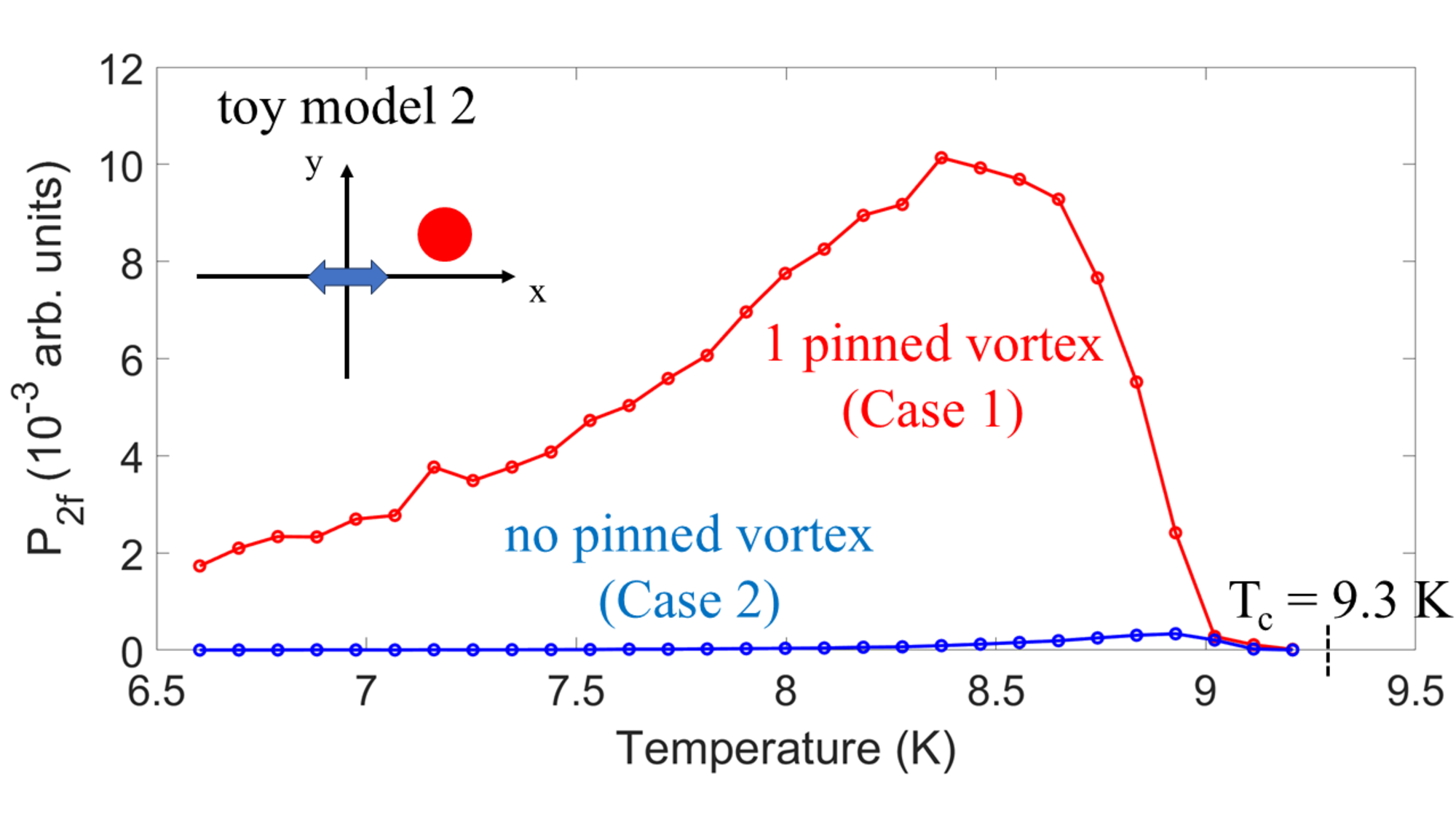}
\caption{\label{fig:TDGLP2fTWithAndWithoutVortex}TDGL simulation results of $P_\mathrm{2f}(T)$ ($B_\mathrm{pk}=28.3$ mT) for toy model 2 under two scenarios. There is one pinned vortex for the red curve (Case 1) and there is no pinned vortex for the blue curve (Case 2). The inset shows the top-view setup of toy model 2: The blue double arrows represent the rf dipole, and the red dot represents the vortex (Case 1) pinned by pinning sites.}
\end{figure}

Core Feature 1: $P_\mathrm{2f}$ arises exclusively in the presence of trapped vortices and is absent when no trapped vortices are present.

Figure \ref{fig:TDGLP2fTWithAndWithoutVortex} shows $P_\mathrm{2f}(T)$ for toy model 2 under two scenarios. In Case 1 (red curve), a single pinned vortex is present, while in Case 2 (blue curve), no pinned vortex is present. Both cases use identical conditions and pinning potentials, with the only difference being the presence or absence of a trapped vortex at the pinning sites. In Case 1, the wiggling of the trapped vortex generates $P_\mathrm{2f}$. In Case 2, there is no trapped vortex, and hence $P_\mathrm{2f}=0$. The slight bump observed in the blue curve around 8.8 K is attributed to numerical errors (slight asymmetries in the computed response between the two half-cycles).


\subsection{Trapped vortex configuration and $P_\mathrm{2f}$}
\label{sec:Feature2}

\begin{figure}
\includegraphics[width=0.45\textwidth]{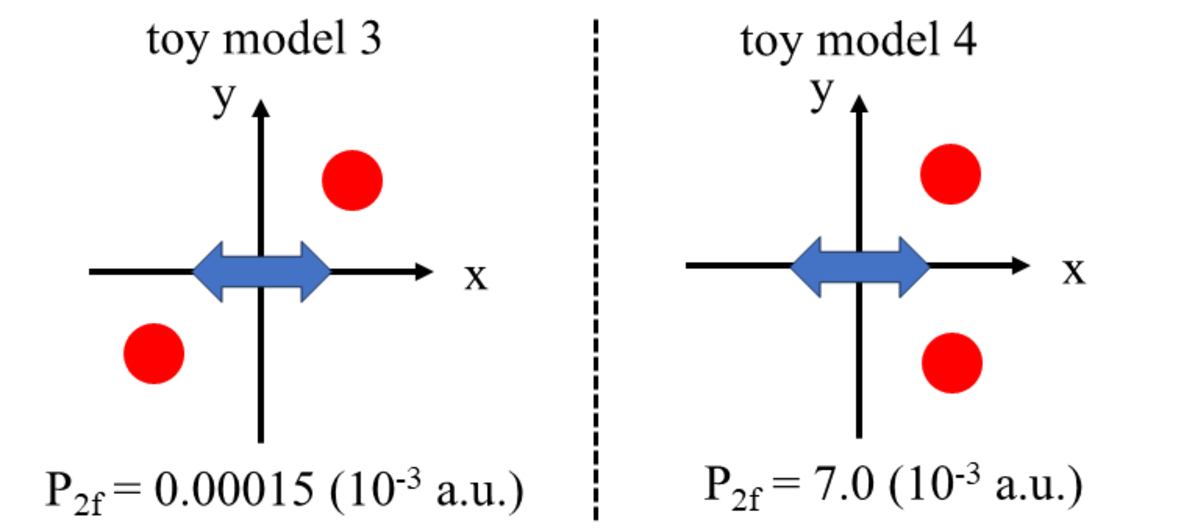}
\caption{\label{fig:TDGLConfiguration}TDGL simulation results of $P_\mathrm{2f}$ for toy models 3 and 4, at $T=7.44$ K and $B_\mathrm{pk}=28.3$ mT. The top-view shows a projection of the rf dipole (blue arrows) and the red dots represent vortices pinned by pinning sites.}
\end{figure}

Core Feature 2: $P_\mathrm{2f}$ is highly sensitive to the specific configuration of trapped vortices.

Figure \ref{fig:TDGLConfiguration} shows $P_\mathrm{2f}$ for toy models 3 and 4, both of which contain two trapped vortices. Despite this similarity, the values of $P_\mathrm{2f}$ differ dramatically. In fact, $P_\mathrm{2f}$ is expected to be zero for toy model 3, as the vortex dynamics is perfectly symmetric by design; the small nonzero value observed in simulations (noted on the Figure) arises from numerical errors. This comparison between toy models 3 and 4 highlights that $P_\mathrm{2f}$ can vary significantly with changes in the trapped vortex configuration. Since $P_\mathrm{2f}$ reflects the time-asymmetry in vortex dynamics, it is inherently sensitive to the detailed arrangement of trapped vortices.


\subsection{Changes in trapped vortex configuration}
\label{sec:Feature3}

\begin{figure}
\includegraphics[width=0.45\textwidth]{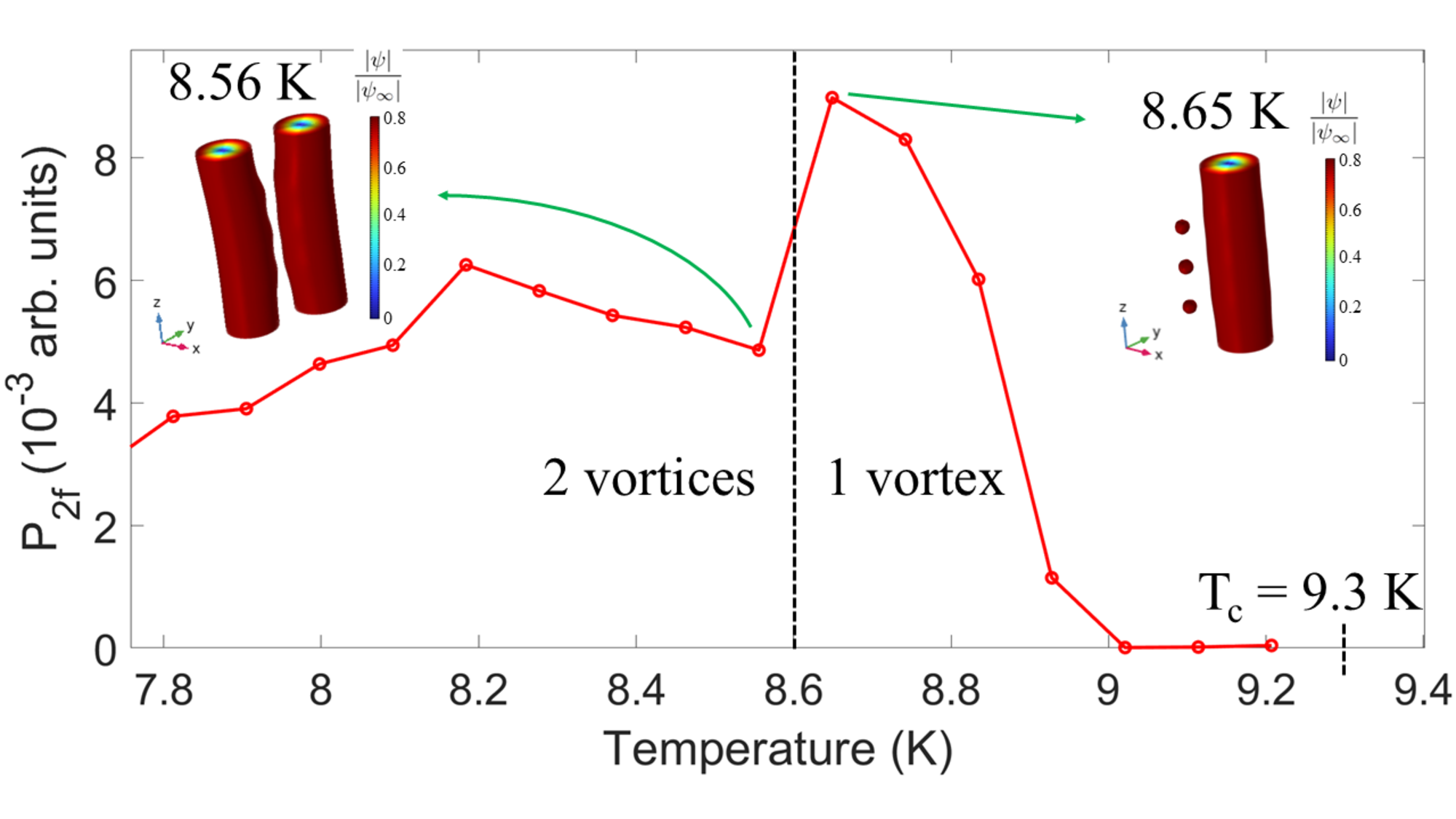}
\caption{\label{fig:TDGLP2fJumpingUp}TDGL simulation results for $P_\mathrm{2f}(T)$ for toy model 5 at $B_\mathrm{pk}=28.3$ mT. The two insets display the normalized order parameter ($\psi / \psi_{\infty}$) at T=8.56 K and T=8.65 K, providing a visualization of the trapped vortices and pinning sites (red spheres).}
\end{figure}

Core Feature 3: Changes in the trapped vortex configuration can cause abrupt jumps in $P_\mathrm{2f}(T)$.

One manifestation of Core Feature 2 is that vortex configuration changes as the temperature increases, which can result in a jump (up or down) in $P_\mathrm{2f}(T)$. Figure \ref{fig:TDGLP2fJumpingUp} shows $P_\mathrm{2f}(T)$ for toy model 5. At low temperatures (T$<$8.6 K), two vortices remain pinned. As the temperature increases beyond 8.6 K, one of the pinned vortices becomes depinned and escapes from the simulation due to the vortex-vortex repulsive force exceeding the pinning force provided by the pinning sites. This depinning event causes a dramatic jump in $P_\mathrm{2f}(T)$ around 8.6 K, as shown in Fig. \ref{fig:TDGLP2fJumpingUp}.

Since $P_\mathrm{2f}$ reflects the time-asymmetry in vortex dynamics rather than being proportional to the number of trapped vortices, its value can either increase or decrease when a pinned vortex is removed. An example of $P_\mathrm{2f}(T)$ exhibiting a downward jump upon vortex depinning is provided in Appendix \ref{sec:P2fTJumpDown}.


\section{Experimental results}
\label{sec:Results}

\subsection{Sample information}
\label{sec:Sample}


\begin{figure}
\includegraphics[width=0.45\textwidth]{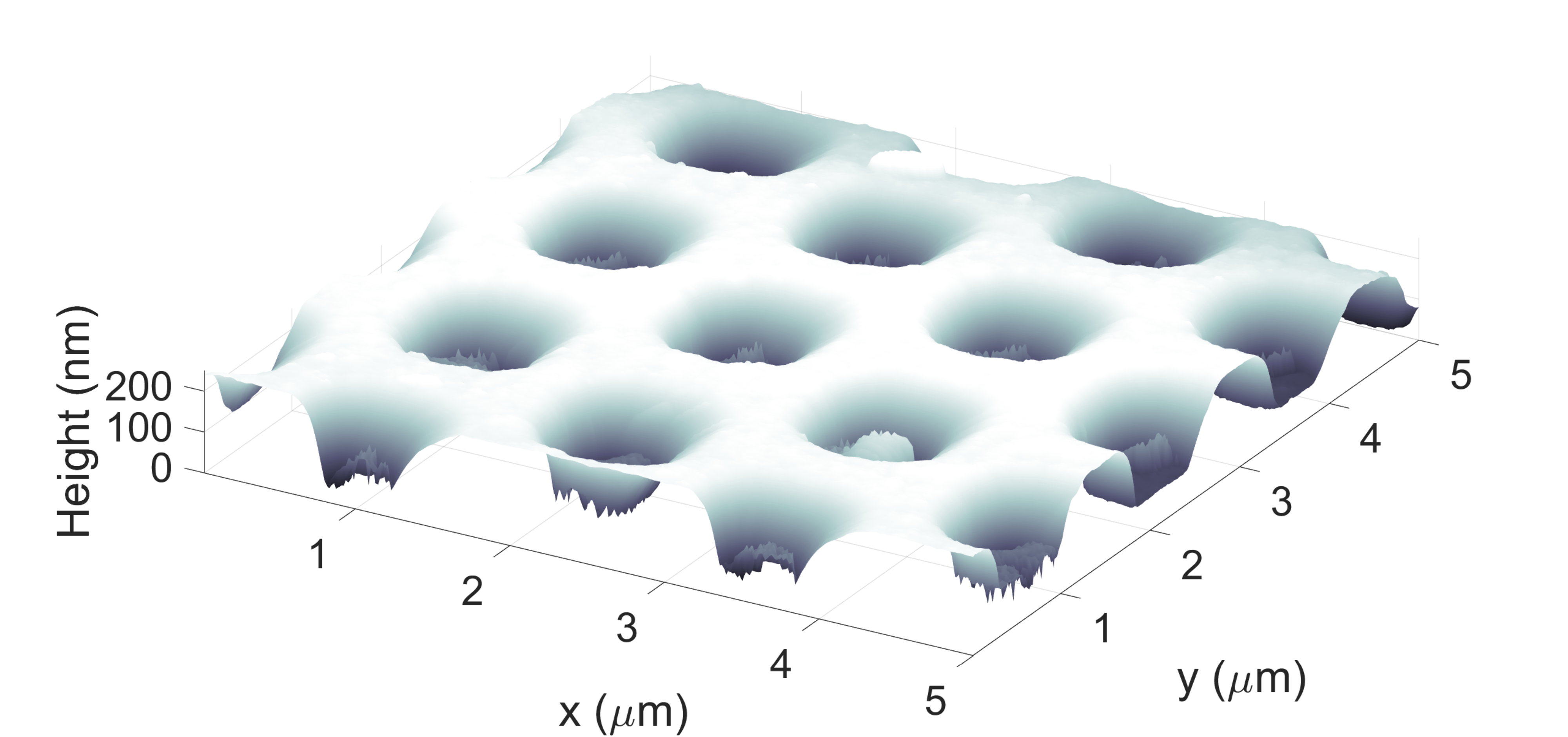}
\caption{\label{fig:SampleTopography}An atomic force microscopy (AFM) image of a $5 \times 5 \, \mu\mathrm{m}^2$ surface area of the Nb antidot film sample.}
\end{figure}

Antidot arrays in superconducting films serve as engineered traps for vortices \cite{Moshchalkov1998Pinning,Raedts2004FluxPinning,Silhanek2005Enhanced,Souza2006Vortex,Berdiyorov2006Novel,Berdiyorov2007Superconducting,Plourde2009Nanostructured,Jin2010High,Cuadra2015rfcoil,Zhang2019Direct}. In this work, we study a 170-nm-thick Nb film with an antidot flux pinning array on a SiOx substrate. The antidot array has a period of $1.6 \, \mu$m and an antidot diameter of $1.2 \, \mu$m. Figure \ref{fig:SampleTopography} shows an atomic force microscopy (AFM) image of a $5 \times 5 \, \mu\mathrm{m}^2$ surface area of the sample. Note that the substrate is exposed in the antidot regions of the film.

\subsection{Measurement protocol}
\label{sec:MeasurementProtocol}

A representative measurement protocol is as follows. Initially, the probe (magnetic writer head) is positioned away from the sample. To create trapped vortices, the sample is first warmed to 10 K, well above the transition temperature. A DC magnetic field ($B_\mathrm{cooldown}$) is then applied, and then the sample is gradually cooled from 10 K (above $T_\mathrm{c}$) to 6.5 K (below $T_\mathrm{c}$) at a rate of 0.55 mK/s. The sample is subsequently cooled rapidly from 6.5 K to 3.8 K (base temperature). Once the temperature stabilizes at 3.8 K, the DC magnetic field is \textit{removed}. The probe is then brought into contact with the sample by adjusting the piezo stages. The microwave source is then activated with a fixed input frequency $f$ and power, and $P_\mathrm{2f}$ is measured as the sample is gradually warmed from 3.8 K to 10 K. During this warmup process, the surface of the sample is exposed to a fixed rf magnetic field, $B_\mathrm{RF}\mathrm{sin}(\omega t)$, over a micron-scale area. Notably, \textit{no} DC magnetic field is applied during the microwave measurements, ensuring that the measured $P_\mathrm{2f}$ originates solely from trapped vortices in the sample.

The magnetic writer head itself exhibits temperature-independent nonlinearity. Since the measured $P_\mathrm{2f}$ is a combination of the probe background and the sample contribution, the probe background, determined by averaging the magnitude of $P_\mathrm{2f}(T)$ between 9.5 K and 10 K, is subtracted from the total signal to isolate the sample response. See Appendix \ref{sec:ProbeBackground} for further details about the background subtraction process.

\subsection{Representative data of $P_\mathrm{2f}(T)$}
\label{sec:RepresentativeData}


\begin{figure}
\includegraphics[width=0.45\textwidth]{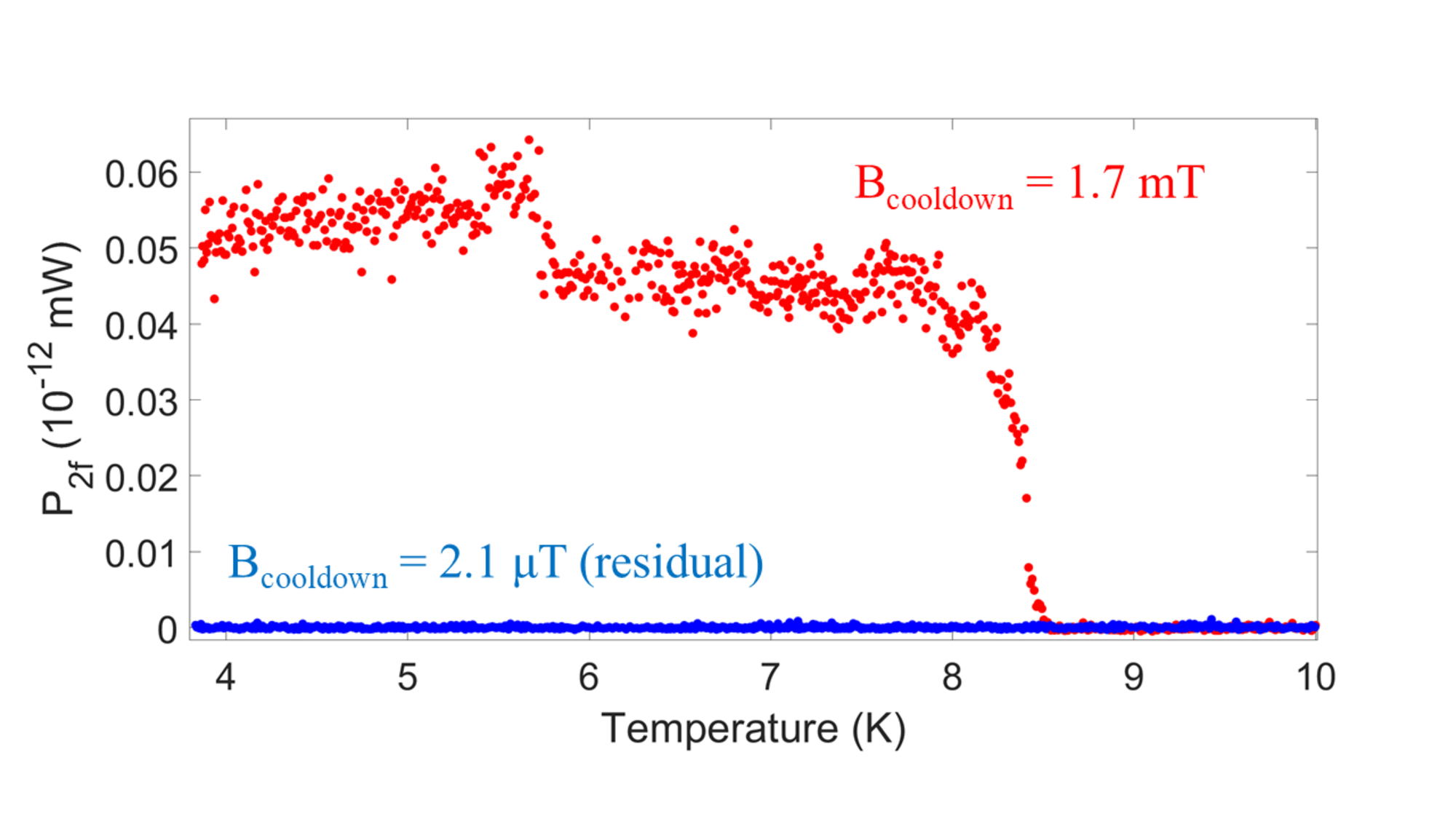}
\caption{\label{fig:P2fWithAndWithoutB}Representative data for $P_\mathrm{2f}$ as a function of temperature during a warmup measurement for two different cooldown fields. The red dots are the data for $B_\mathrm{cooldown} = 1.7 $ mT and the blue dots are the data for $B_\mathrm{cooldown} = 2.1 \, \mu$T. The input frequency is 1.81 GHz, and the input power is 6.8 dBm.}
\end{figure}

Figure \ref{fig:P2fWithAndWithoutB} shows the representative data for the linear power scale $P_\mathrm{2f}(T)$ during a warmup measurement at a fixed location on the sample, for two different cooldown fields. The input frequency is $f = \omega/2\pi = 1.81$ GHz, and the input power is +6.8 dBm (at the room temperature microwave source).

The red dots are the data for $B_\mathrm{cooldown} = 1.7$ mT. This cooldown magnetic field corresponds to an average flux quanta separation ($\sqrt{\Phi_0 / B_\mathrm{cooldown}}$) of $1.11 \, \mu$m, a filling fraction of the antidot lattice of 1.70, and an average of 8.3 flux quanta in the field of view of the microwave microscope probe (see Appendix \ref{sec:SpatialResolution}). The $P_\mathrm{2f}(T)$ shows a clear transition around 8.6 K, which is the $T_\mathrm{c}$ of the sample. For T$<$8.6 K, vortices are trapped in the sample, and their wiggling under stimulation by the rf magnetic field generates $P_\mathrm{2f}$. For T$>$8.6 K, the sample loses superconductivity and the trapped vortices are released, and hence $P_\mathrm{2f}$ vanishes. 

The blue dots are the data for $B_\mathrm{cooldown} = 2.1 \, \mu$T, which is the measured residual DC magnetic field when no external field is applied. This cooldown magnetic field corresponds to an average flux quanta separation of $31.40 \, \mu$m, a filling fraction of the antidot lattice of 0.0021, and an average of 0.01 flux quanta in the field of view of the microwave microscope probe. Because the density of trapped vortices is very low, with high probability, there are no trapped vortices in the field of view of the probe, and thus there is no $P_\mathrm{2f}$.

The contrast between the two data sets (red and blue) in Fig. \ref{fig:P2fWithAndWithoutB} clearly demonstrates that $P_\mathrm{2f}$ originates from vortices trapped in the sample that are in the field of view of the probe, which agrees with Core Feature 1 (see Sec. \ref{sec:Feature1}).

For simplicity, in the following text, the term ``trapped vortices" will refer specifically to those within the field of view of the microwave microscope probe.


A distinct discrete jump in $P_\mathrm{2f}(T)$ is observed around 5.75 K for $B_\mathrm{cooldown} = 1.7 $ mT in Fig. \ref{fig:P2fWithAndWithoutB}. In the following, the temperature of such a discrete jump is denoted as $T_{P_\mathrm{2f}(T)\mathrm{jump}}$. One possible explanation for this $P_\mathrm{2f}(T)$ jump is as follows: For temperatures slightly below 5.75 K, the trapped vortices are arranged in a specific configuration. As the temperature increases to approximately 5.75 K, the configuration of the trapped vortices changes. This change in configuration leads to the discrete jump in $P_\mathrm{2f}(T)$ (see Sec. \ref{sec:Feature3}), since the measured $P_\mathrm{2f}$ comes from the superposition of the contributions of all the trapped vortices under the probe. The interpretation of $P_\mathrm{2f}(T)$ jumps as indicators of vortex configuration changes is supported by the observation of $P_\mathrm{2f}$ hysteresis in temperature sweeps (see Appendix \ref{sec:HysteresisCheck}).

\subsection{Statistics of $P_\mathrm{2f}(T)$ jumps}
\label{sec:JumpStatistics}


\begin{figure}
\includegraphics[width=0.45\textwidth]{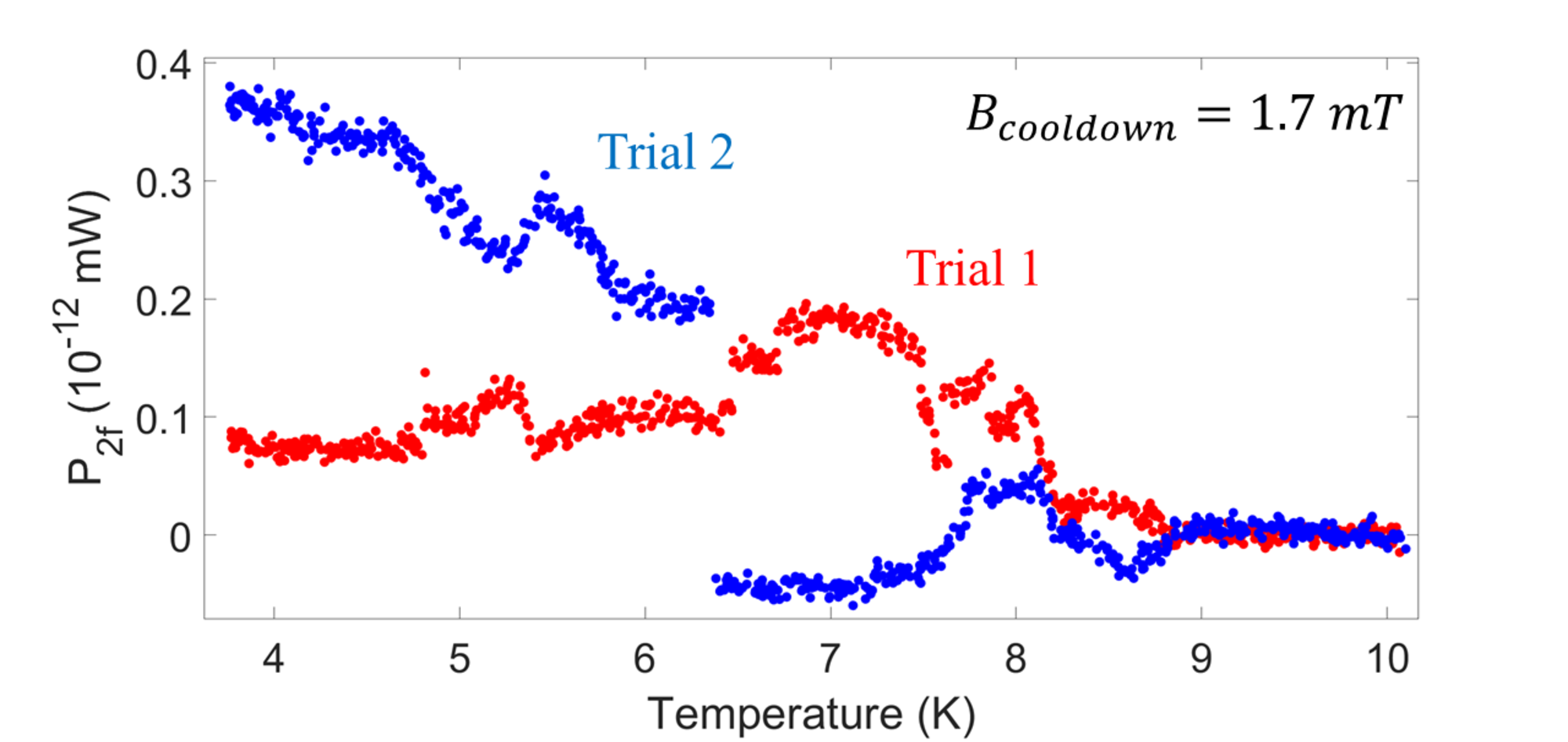}
\caption{\label{fig:P2fStochastic}Two repeated measurements of $P_\mathrm{2f}(T)$ at nominally the same location of the Nb film sample for $B_\mathrm{cooldown} = 1.7 $ mT. The input frequency is 1.819 GHz, and the input power is -5 dBm.}
\end{figure}

The measured value of $P_\mathrm{2f}$ is highly sensitive to the details of trapped vortex configurations (see Sec. \ref{sec:Feature2}), which are complex and inherently stochastic. For example, Fig. \ref{fig:P2fStochastic} presents two repeated measurements of $P_\mathrm{2f}(T)$ performed at the same location under nominally identical conditions, where the contrast between Trial 1 (red) and Trial 2 (blue) highlights this stochastic behavior. (A discussion on the negative $P_\mathrm{2f}(T)$ observed in Trial 2 is provided in Appendix \ref{sec:ProbeBackground}.) Given this sensitivity, rather than focusing on the absolute value of $P_\mathrm{2f}$, we analyze $P_\mathrm{2f}(T)$ jumps, which serve as indicators of changes in trapped vortex configurations and are therefore closely related to pinning properties.

To further investigate $P_\mathrm{2f}(T)$ jumps, we perform repeated measurements of $P_\mathrm{2f}(T)$ from 3.8 K to above $T_{c}$ at nominally the same location of the Nb film sample and analyze the statistical properties of $T_{P_\mathrm{2f}(T)\mathrm{jump}}$.

\begin{figure}
\includegraphics[width=0.45\textwidth]{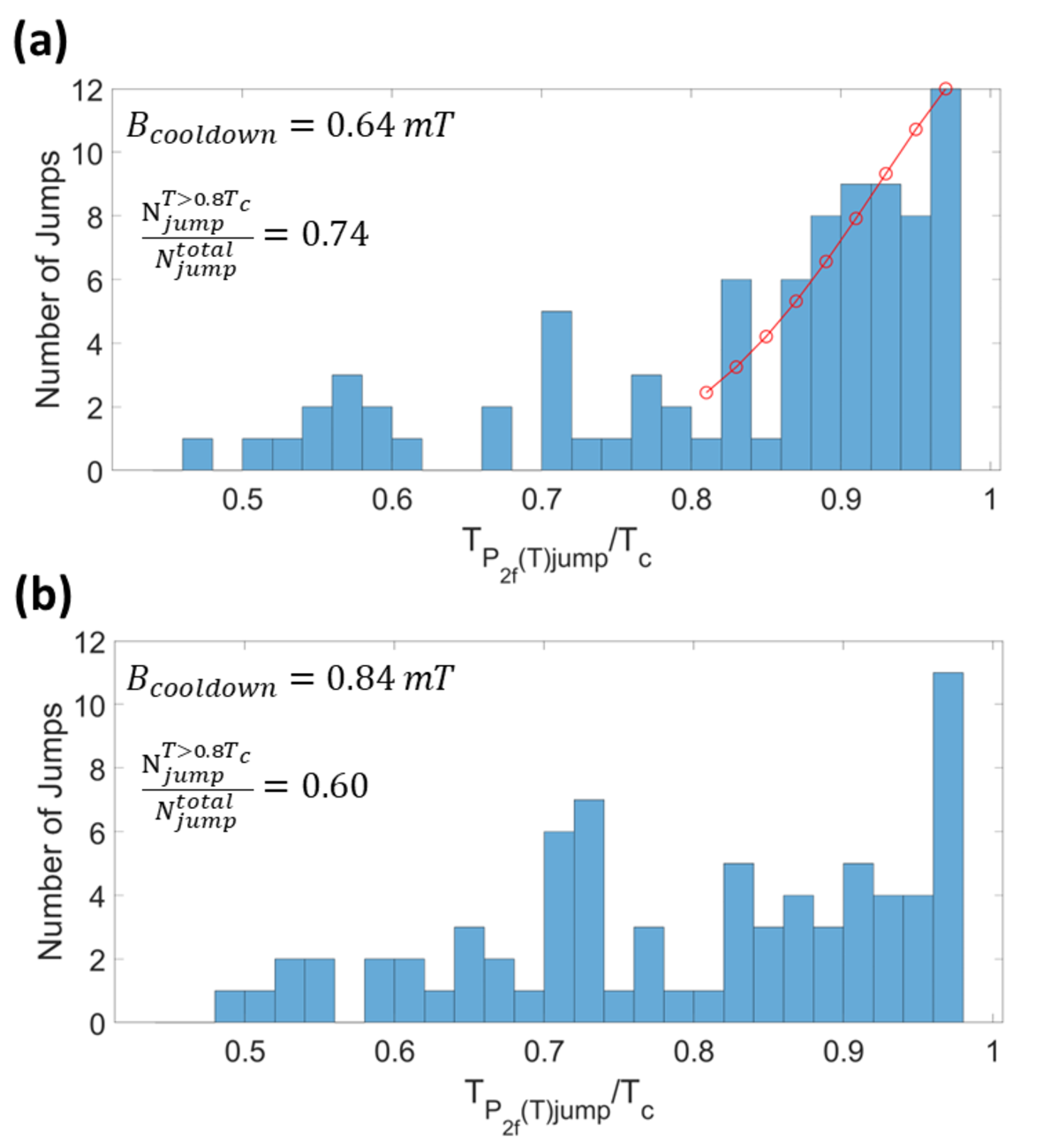}
\caption{\label{fig:P2fJumpHistogram2B}Histogram of the temperatures at which $P_\mathrm{2f}(T)$ jumps occur ($T_{P_\mathrm{2f}(T)\mathrm{jump}}$). The data is obtained from 10 repeated measurements of $P_\mathrm{2f}(T)$ from 3.8 K to above $T_{c}$ at Location 1. Panels (a) and (b) correspond to cooldown fields of $B_\mathrm{cooldown} = 0.64$ mT and $B_\mathrm{cooldown} = 0.84$ mT, respectively. The input frequency is 1.708 GHz, and the input power is -3 dBm. The red curve in panel (a) is the fitting result using Eqs.~\ref{eq:PinningPotentialFormula} and \ref{eq:JumpProbabilityFormula}.}
\end{figure}

Figure \ref{fig:P2fJumpHistogram2B}(a) presents the statistical distribution of $T_{P_\mathrm{2f}(T)\mathrm{jump}}$ as a function of temperature. The data is obtained from 10 repeated measurements of $P_\mathrm{2f}(T)$ during a temperature sweep from 3.8 K to above $T_{c}$, conducted nominally at the same location on the sample surface, with $B_\mathrm{cooldown} = 0.64$ mT. This cooldown magnetic field corresponds to an average flux quanta separation of $1.80 \, \mu$m, a filling fraction of the antidot lattice of 0.65, and an average of 3.2 flux quanta in the field of view of the microwave microscope probe. 

The distribution in Fig. \ref{fig:P2fJumpHistogram2B}(a) reveals three distinct clusters: one above $T_{P_\mathrm{2f}(T)\mathrm{jump}}/T_\mathrm{c} = 0.8$, another peaked around $T_{P_\mathrm{2f}(T)\mathrm{jump}}/T_\mathrm{c} = 0.72$, and a third below $T_{P_\mathrm{2f}(T)\mathrm{jump}}/T_\mathrm{c} = 0.6$.

The rearrangement of trapped vortices as the temperature increases is inherently complex, potentially resulting in a multi-component $T_{P_\mathrm{2f}(T)\mathrm{jump}}$ distribution. For example, an Abrikosov vortex might jump from one pinning site to a nearby pinning site \cite{Sanders1993Thermally,Chen2024Revealing}, be attracted to and captured by an antidot, or flux trapped by antidots could rearrange into other antidots near $T_{c}$.

The three distinct clusters in Fig. \ref{fig:P2fJumpHistogram2B}(a) suggest the presence of three different types of trapped vortex rearrangement events in the sample. The profiles and temperatures of these clusters provide insight into the pinning properties of the sample. Notably, the cluster above $T_{P_\mathrm{2f}(T)\mathrm{jump}}/T_\mathrm{c} = 0.8$ accounts for more than half of the total $P_\mathrm{2f}(T)$ jumps. Specifically, the high-temperature jump ratio, defined as $N^{T>0.8T_{c}}_\mathrm{jump}/N^\mathrm{total}_\mathrm{jump}$, is 0.74 for Fig. \ref{fig:P2fJumpHistogram2B}(a).

One possible mechanism for trapped vortex rearrangement events is thermal activation \cite{Beasley1969Flux,Coffey1991Unified,Sanders1993Thermally,Sok1994Thermal,Golosovsky1996High,Gail1998Pinning,Enrico2017Vortices}. Under this mechanism, trapped vortices are expected to rearrange more frequently at higher temperatures, consistent with the trend observed in the cluster above $T_{P_\mathrm{2f}(T)\mathrm{jump}}/T_\mathrm{c} = 0.8$ in Fig. \ref{fig:P2fJumpHistogram2B}(a). To analyze the data further, assume a temperature-dependent activation energy $U(T)$ as \cite{Yeshurun1988Giant,Tinkham1988Resistive,Gail1998Pinning,Mart2000Temperature,Zhang2006Thermally,Zhang2009Thermally,Choi2017Thermal,Altanany2024Vortex}
\begin{equation} \label{eq:PinningPotentialFormula}
U(T)=U_0 \left( 1-\frac{T}{T_\mathrm{c}} \right) ^\gamma
\end{equation}
where $\gamma$ is a scaling exponent. Assume the likelihood of trapped vortex rearrangement events $P(T)$ due to thermal activation at temperature $T$ is given by
\begin{equation} \label{eq:JumpProbabilityFormula}
P(T) \propto e^{-\frac{U(T)}{k_{B}T}}.
\end{equation}
Assuming that trapped vortex rearrangement events for $T > 0.8~T_\mathrm{c}$ in Fig. \ref{fig:P2fJumpHistogram2B}(a) are primarily driven by thermal activation, with vortex-vortex interactions playing a minor role, we fit the probability distribution above $T = 0.8~T_\mathrm{c}$ using Eqs.~\ref{eq:PinningPotentialFormula} and \ref{eq:JumpProbabilityFormula}. The resulting fit yields a pinning potential of $U_0 = 14.8~k_B T_\mathrm{c} = 127.3~\mathrm{K} = 10.97~\mathrm{meV}$ and a scaling exponent of $\gamma = 1.43$.


\begin{table}
\begin{center}
\begin{tabular}{ | m{1.4cm} | m{1.4cm}| m{1.2cm} | m{3.2cm} | }
\hline
Location & $U_0~(\mathrm{meV})$ & $\gamma$ & $U(T)~(\mathrm{meV})$ for $0.30<T/T_\mathrm{c}<0.48$ \\ 
\hline
1 & 10.97 & 1.43 & between 4.26 and 6.57 \\ 
\hline
2 & 9.86 & 1.76 & between 3.08 and 5.25 \\ 
\hline
\end{tabular}
\caption{Summary of the vortex pinning parameters ($U_0$ and $\gamma$) extracted from $P_\mathrm{2f}(T)$ measurements using Eqs.~\ref{eq:PinningPotentialFormula} and \ref{eq:JumpProbabilityFormula}. Data are presented for two distinct locations on the sample surface (Location 1 and Location 2), separated by 50~$\mu$m.}
\label{tbl:PinningParameter}
\end{center}
\end{table}

Building on this analysis, we investigate the spatially-resolved vortex pinning properties ($U_0$ and $\gamma$) of the sample. We perform measurements following the same procedure as described in Fig. \ref{fig:P2fJumpHistogram2B}(a) (Location 1) but at a different location on the sample surface (Location 2). Locations 1 and 2 are separated by 50 $\mu$m. The extracted pinning parameters for these two locations are summarized in Table~\ref{tbl:PinningParameter}.

Several previous studies have described the temperature dependence of the activation energy $U(T)$ using the form given by Eq.~\ref{eq:PinningPotentialFormula}. Typically, the exponent $\gamma$ falls within the range $0.5 < \gamma < 2$ \cite{Zhang2006Thermally,Zhang2009Thermally,Choi2017Thermal}. In our analysis, we obtained $\gamma = 1.43$ and $1.76$ at Locations 1 and 2, respectively, which are consistent with values commonly reported in the literature.

Vortex pinning properties in Nb films have previously been studied in Ref.~\cite{Gail1998Pinning}, where a temperature-independent activation energy of $1.9~\mathrm{meV}$ was obtained by fitting experimental data in the temperature range $0.3 < T/T_\mathrm{c} < 0.48$. In contrast, we use a temperature-dependent activation energy $U(T)$ described by Eq.~\ref{eq:PinningPotentialFormula}; our fitted values in the same temperature range, summarized in the last column of Table~\ref{tbl:PinningParameter}, are comparable to that reported in Ref.~\cite{Gail1998Pinning}. Furthermore, our results also agree in magnitude with the activation energies of Pb-Tl alloy, reported to be between $3~\mathrm{meV}$ and $14~\mathrm{meV}$ in Ref.~\cite{Beasley1969Flux}. 

Taken together, these comparisons confirm that the pinning parameters ($U_0$ and $\gamma$) obtained from our analysis are consistent with values reported in the literature, thus validating the physical reasonability of our results.


Figure \ref{fig:P2fJumpHistogram2B}(b) shows the statistical distribution of $T_{P_\mathrm{2f}(T)\mathrm{jump}}$ as a function of temperature for $B_\mathrm{cooldown} = 0.84$ mT. The measurement procedure is identical to that of Fig. \ref{fig:P2fJumpHistogram2B}(a), except with a higher $B_\mathrm{cooldown}$, resulting in an increased trapped vortex density. Compared to Fig. \ref{fig:P2fJumpHistogram2B}(a) ($N^{T>0.8T_{c}}_\mathrm{jump}/N^\mathrm{total}_\mathrm{jump} = 0.74$), the distribution of $T_{P_\mathrm{2f}(T)\mathrm{jump}}$ in Fig. \ref{fig:P2fJumpHistogram2B}(b) is more spread out, with a lower high-temperature jump ratio of $N^{T>0.8T_{c}}_\mathrm{jump}/N^\mathrm{total}_\mathrm{jump} = 0.60$.


\begin{table}
\begin{center}
\begin{tabular}{ | m{1.8cm} | m{1.8cm}| m{1.8cm} | m{1.8cm} | }
\hline
Cooldown from room temperature to 3.8 K & Location & input frequency (GHz) & input power (dBm) \\ 
\hline
\multirow{2}{*}{First} & 1 & \multirow{2}{*}{1.708} & \multirow{2}{*}{-3} \\
& 2 & & \\ 
\hline
\multirow{2}{*}{Second} & 3 & \multirow{2}{*}{1.819} & \multirow{2}{*}{-5} \\
& 4 & & \\ 
\hline
\end{tabular}
\caption{Summary of the measurement details at four distinct locations on the sample surface. Locations 1 and 2 are separated by 50 $\mu$m, as are Locations 3 and 4. Measurements at Locations 1 and 2 were performed during the same cooldown from room temperature to 3.8 K, while measurements at Locations 3 and 4 were obtained in a separate cooldown from room temperature to 3.8 K.}
\label{tbl:FourLocationDetails}
\end{center}
\end{table}

\begin{figure}
\includegraphics[width=0.45\textwidth]{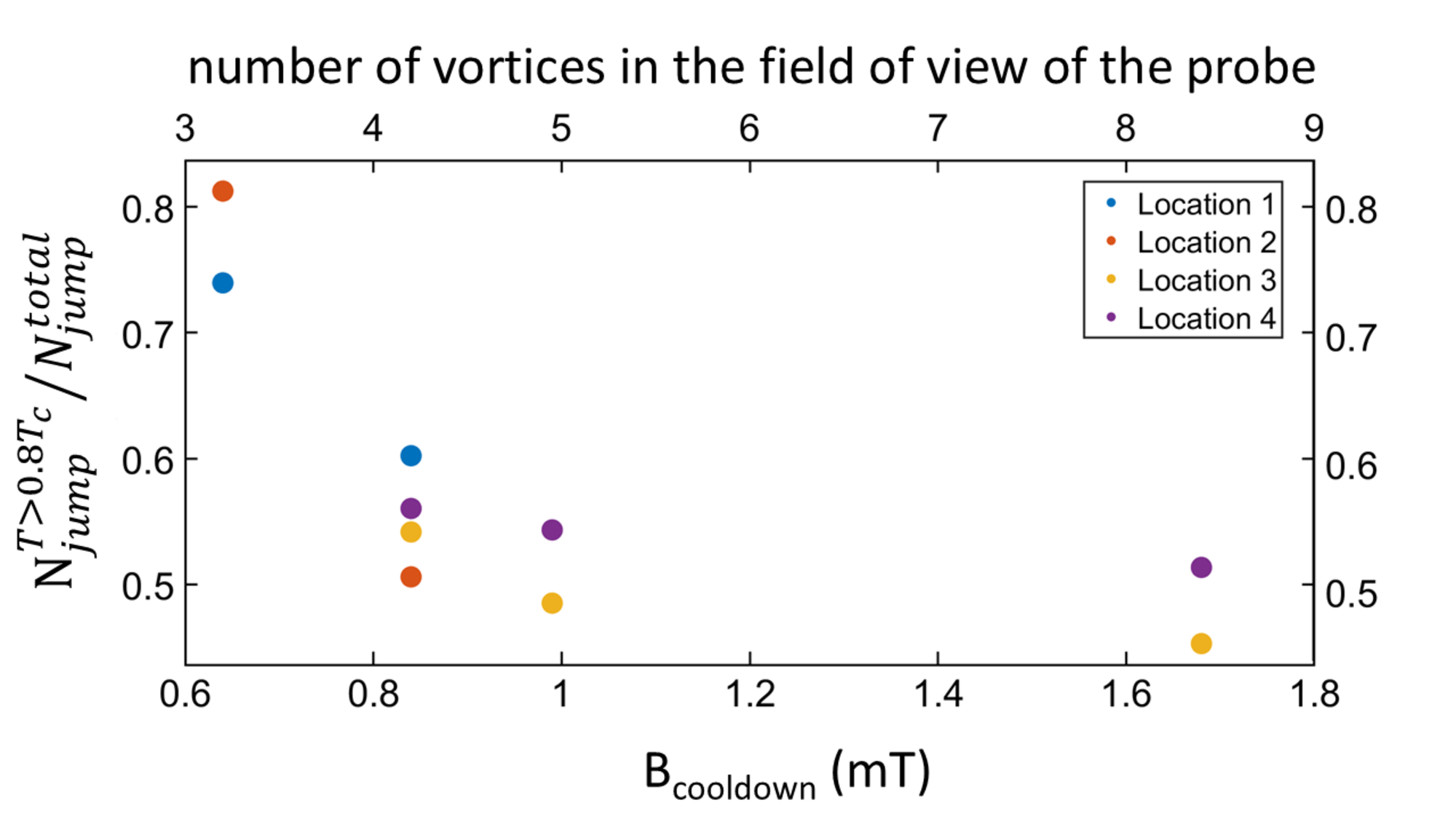}
\caption{\label{fig:P2fJumpHighTRatio}High-temperature jump ratio $N^{T>0.8T_{c}}_\mathrm{jump}/N^\mathrm{total}_\mathrm{jump}$ for different $B_\mathrm{cooldown}$ values, measured at four distinct locations on the sample surface. The measurement details at the four locations are summarized in Table \ref{tbl:FourLocationDetails}. The data in Fig. \ref{fig:P2fJumpHistogram2B} is obtained at Location 1.}
\end{figure}

Building on the results from Fig. \ref{fig:P2fJumpHistogram2B}, we extend the measurements to multiple $B_\mathrm{cooldown}$ values, corresponding to different trapped vortex densities, at four distinct locations on the sample surface (Locations 1 to 4). The measurement details at the four locations are summarized in Table \ref{tbl:FourLocationDetails}. Figure \ref{fig:P2fJumpHighTRatio} presents the high-temperature jump ratio $N^{T>0.8T_{c}}_\mathrm{jump}/N^\mathrm{total}_\mathrm{jump}$ as a function of $B_\mathrm{cooldown}$. 

As shown in Fig. \ref{fig:P2fJumpHighTRatio}, $N^{T>0.8T_{c}}_\mathrm{jump}/N^\mathrm{total}_\mathrm{jump}$ decreases with increasing $B_\mathrm{cooldown}$ across all four measurement locations. One possible explanation for this trend is as follows. In the low trapped vortex density regime (small $B_\mathrm{cooldown}$), vortex-vortex interactions are weak, and trapped vortex rearrangement events are primarily driven by thermal activation, which is most significant near $T_\mathrm{c}$. In contrast, in the high trapped vortex density regime (large $B_\mathrm{cooldown}$), the trapped vortex configuration becomes more complex, as both the pinning potential of the antidot array and vortex-vortex interactions play a significant role. In this regime, trapped vortices can rearrange even at lower temperatures.  

Understanding trapped vortex rearrangement mechanisms beyond thermal activation near $T_\mathrm{c}$ requires more sophisticated theoretical analysis and numerical simulations, which are beyond the scope of this work.


\section{Conclusion}
\label{sec:Conclusion}

In this work, we present a proof-of-principle study demonstrating the use of local $P_\mathrm{2f}$ measurements with our near-field magnetic microwave microscope to investigate the dynamics of trapped vortices at the micron scale. This microscopic approach complements traditional methods that average the behavior of many vortices. Specifically, we show that this technique can reveal depinning events and provide insights into spatially-resolved pinning properties. Furthermore, our results highlight how vortex pinning behavior varies across different trapped vortex density regimes, offering a new perspective on the interplay between pinning potentials and vortex-vortex interactions. It is particularly relevant for applications such as SRF cavities, superconducting quantum circuits, and superconducting digital electronics. Future work includes developing more sophisticated models beyond the thermal activation mechanism to further extract quantitative information about pinning properties from $P_\mathrm{2f}$ measurements.


\section{Acknowledgement}
\label{sec:Acknowledgement}

We would like to thank Javier Guzman from Seagate Technology for providing magnetic write heads, and Cougar Garcia for assistance. We also thank Thomas Antonsen for helpful discussions. This work is supported by DOE/HEP under grant DESC0017931, and by ARO/FSDL under grant ARO W911NF-24-1-0153.

\appendix


\section{Field of view of the near-field microwave probe}
\label{sec:SpatialResolution}

The field of view of the probe can be estimated using a Monte Carlo approach. For instance, if the average number of trapped vortices within the field of view of the probe for a given $B_\mathrm{cooldown}$ value is 0.4, the probability of the probe interacting with a trapped vortex and detecting the $P_\mathrm{2f}$ signal is 40\%. In practice, we perform 6 repeated measurements of $P_\mathrm{2f}(T)$ (with an input frequency of 1.824 GHz and input power of 6 dBm) and record how many trials exhibit a transition near the $T_\mathrm{c}$ of the sample in the low trapped vortex density regime (small $B_\mathrm{cooldown}$ values). For $B_\mathrm{cooldown} =$ 0.061 mT, 0.124 mT, and 0.188 mT, the number of trials showing a sample signal out of 6 are 2, 4, and 5, respectively. Based on these results, the interaction range between the probe and trapped vortices is estimated to be approximately $1.1 \, \mu$m.


\section{Equations and parameters for TDGL simulations}
\label{sec:TDGLsetting}

The two TDGL equations solved in this work are \cite{tinkham2004introduction,gulian2020shortcut}

\begin{equation} \label{eq:FirstTDGL}
\begin{split}
\gamma\frac{\partial\psi}{\partial t}=\frac{-1}{2m_{*}}\left(-i\hbar\nabla-e_{*}A\right)^{2}\psi-\alpha\psi-\beta\left|\psi\right|^{2}\psi
\end{split}
\end{equation}
and
\begin{equation} \label{eq:SecondTDGL}
\begin{split}
\sigma\frac{\partial A}{\partial t}= & \frac{-i\hbar e_{*}}{2m_{*}}\left(\psi^{*}\nabla\psi-\psi\nabla\psi^{*}\right)-\frac{e_{*}^{2}}{m_{*}}\left|\psi\right|^{2}A \\
& -\frac{1}{\mu_{0}}\nabla\times\nabla\times A.
\end{split}
\end{equation}
Here $e_{*}=2e$ is the charge of the Cooper pair, $m_{*}=2m_{e}$ is the mass of the Cooper pair, $\gamma=\frac{\pi \hbar \left|\alpha\right|}{8 k_{B} (T_\mathrm{c}-T)}$ plays the role of a friction coefficient, $\psi$ is the order parameter, $A$ is the total vector potential (arising from both external and self-generated sources), $\sigma$ is the electric conductivity of the normal state, and both $\alpha$ and $\beta$ are material-specific phenomenological parameters. Note that $\alpha$ and $\beta$ can be related to the penetration depth $\lambda$ and the thermodynamic critical field $B_\mathrm{c}$ by

\begin{equation} \label{eq:LambdaFormula}
\lambda=\sqrt{\frac{m_{*}\beta}{\mu_{0}e_{*}^{2}\left|\alpha\right|}}
\end{equation}
and
\begin{equation} \label{eq:BcFormula}
B_\mathrm{c}=\frac{\sqrt{\mu_{0}}\left|\alpha\right|}{\sqrt{\beta}}.
\end{equation}

The total current is the combination of the supercurrent $J_\mathrm{s}$ and the normal current $J_\mathrm{n}$. The supercurrent $J_\mathrm{s}$ is given by
\begin{equation} \label{eq:Js}
J_\mathrm{s} = \frac{-i\hbar e_{*}}{2m_{*}}\left(\psi^{*}\nabla\psi-\psi\nabla\psi^{*}\right)-\frac{e_{*}^{2}}{m_{*}}\left|\psi\right|^{2}A
\end{equation}
and the normal current $J_\mathrm{n}$ is given by
\begin{equation} \label{eq:Jn}
J_\mathrm{n} = \sigma E = -\sigma\frac{\partial A}{\partial t}.
\end{equation}

For boundary conditions, any current passing through the boundary between a superconductor and a vacuum is unphysical. Therefore, the boundary conditions must be enforced along the entire enclosing boundary $\partial \Omega$ of the superconducting simulation domain, including the sides and bottom surfaces. Specifically, we expect:

\begin{equation} \label{eq:BCfirst}
\nabla \psi \cdot \hat{n} = 0 \quad \textrm{on} \quad \partial \Omega.
\end{equation}
\begin{equation} \label{eq:BCsecond}
A \cdot \hat{n} = 0 \quad \textrm{on} \quad \partial \Omega.
\end{equation}
Here $\hat{n}$ is the unit vector normal to the boundary.

In the toy models studied in this work (Niobium superconductor hosting vortex pinning sites), a pinning site is modeled as a spherical region of radius $r_\mathrm{pin}$ composed of a low-$T_\mathrm{c}$ impurity phase. Pinning sites are incorporated by introducing spatial variations in five material-specific parameters: $T_\mathrm{c}$, $\alpha$, $\beta$, $\gamma$, and $\sigma$. The values of $\alpha$ and $\beta$ are determined for a given choice of $\lambda$ and $B_\mathrm{c}$ using Eqs. \ref{eq:LambdaFormula} and \ref{eq:BcFormula}. For simplicity, the normal state conductivity $\sigma$ of the low-$T_\mathrm{c}$ impurity phase is set to be the same as that of Nb. The critical temperature, penetration depth, and critical field of Nb are 9.3 K, 40 nm, and 200 mT, respectively. For the low-$T_\mathrm{c}$ impurity phase, these values are 3.0 K, 90 nm, and 120 mT, respectively.

The location of a pinning site is represented by $(x_\mathrm{pin}, y_\mathrm{pin}, z_\mathrm{pin})$, where $z_\mathrm{pin} < 0$ since the sample occupies the $z < 0$ region. All toy models include either one or two sets of pinning sites, with each set consisting of three identical pinning sites sharing the same $(x_\mathrm{pin}, y_\mathrm{pin})$ coordinates. The $z_\mathrm{pin}$ values for the three sites are -200 nm, -360 nm, and -520 nm. Each set of pinning sites can then be labeled by its $x_\mathrm{pin}$, $y_\mathrm{pin}$, and $r_\mathrm{pin}$. The pinning site configurations are summarized in Table \ref{tbl:ToyModelSetting}.

\begin{table}
\begin{center}
\begin{tabular}{ | m{1.8cm} | m{1.8cm}| m{1.8cm} | m{1.8cm} | }
\hline
toy model & $x_\mathrm{pin}$ (nm) & $y_\mathrm{pin}$ (nm) & $r_\mathrm{pin}$ (nm) \\ 

\hline
1 & 0 & 0 & 40 \\

\hline
2 & 120 & 80 & 40 \\

\hline
\multirow{2}{*}{3} & 80 & 80 & 40 \\
& -80 & -80 & 40 \\ 

\hline
\multirow{2}{*}{4} & 80 & 80 & 40 \\
& 80 & -80 & 40 \\ 

\hline
\multirow{2}{*}{5} & 120 & 80 & 40 \\
& 40 & -120 & 16 \\ 

\hline
\multirow{2}{*}{6} & 120 & 80 & 40 \\
& 120 & -80 & 18 \\ 

\hline
\end{tabular}
\caption{Summary of the pinning site configurations for toy models 1 to 6. All toy models include either one or two sets of pinning sites, with each set consisting of three identical pinning sites sharing the same $(x_\mathrm{pin}, y_\mathrm{pin})$ coordinates. The $z_\mathrm{pin}$ values for the three sites are -200 nm, -360 nm, and -520 nm. See Appendix \ref{sec:P2fTJumpDown} for the discussion of toy model 6.}
\label{tbl:ToyModelSetting}
\end{center}
\end{table}


\section{Magnetic field of a point dipole}
\label{sec:DipoleBField}

Here we present the magnetic field generated by a point dipole in free space at the xy-plane ($z=0$). Consider a pointlike magnetic dipole located at $(x_\mathrm{dp},y_\mathrm{dp},h_\mathrm{dp})$, oriented in the x-direction, with a magnetic moment $M$. The x-component of the dipole magnetic field at $(x,y,z=0)$ is given by
\begin{equation} \label{eq:DipoleB}
B_\mathrm{dp,x}(x,y) = \frac{\mu_0 M}{4 \pi} \frac{3(x-x_\mathrm{dp})^2-r^2}{r^5}, 
\end{equation}
where $r=\sqrt{(x-x_\mathrm{dp})^2+(y-y_\mathrm{dp})^2+h_\mathrm{dp}^2}$. For $B_\mathrm{dp,x} \nabla_{x} B_\mathrm{dp,x}$, we have
\begin{equation} \label{eq:DipoleBGradB}
\begin{split}
B_\mathrm{dp,x} \nabla_{x} B_\mathrm{dp,x} = & \left( \frac{\mu_0 M}{4 \pi}\right)^2 \frac{3(x-x_\mathrm{dp})^2-r^2}{r^{12}} \\
& \times 3(x-x_\mathrm{dp})\left[3r^2-5(x-x_\mathrm{dp})^2\right].
\end{split}
\end{equation}


\section{Another simulated $P_\mathrm{2f}(T)$ for two pinned vortices}
\label{sec:P2fTJumpDown}

\begin{figure}
\includegraphics[width=0.45\textwidth]{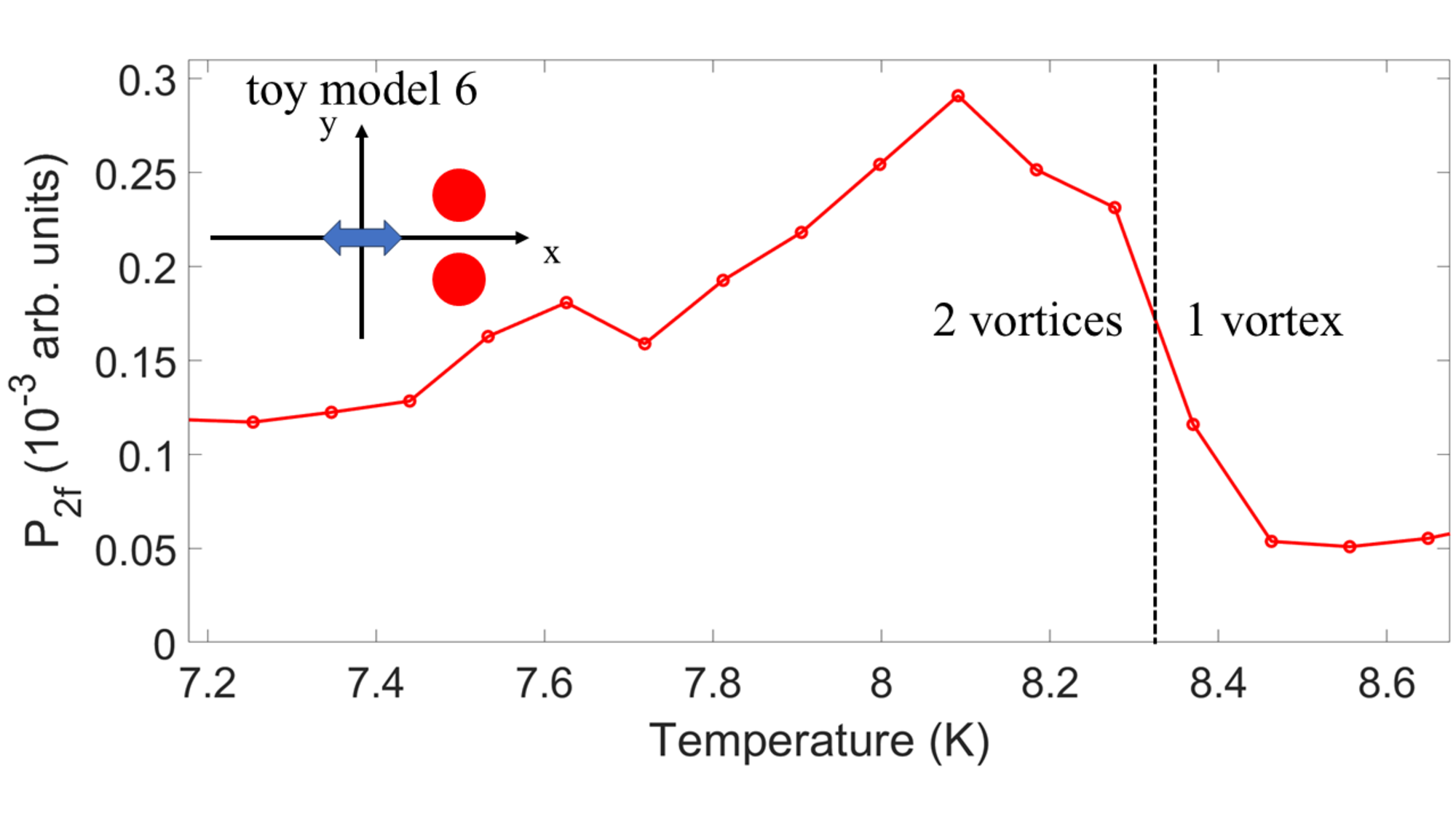}
\caption{\label{fig:TDGLP2fJumpingDown}TDGL simulation results for $P_\mathrm{2f}(T)$ for toy model 6 at $B_\mathrm{pk}=5.7$ mT. The blue double arrow represents the rf dipole, and the red dots represent vortices pinned by pinning sites.}
\end{figure}

Here we provide another example of simulated $P_\mathrm{2f}(T)$. Figure \ref{fig:TDGLP2fJumpingDown} shows $P_\mathrm{2f}(T)$ for toy model 6. The setting of pinning sites configuration of toy model 6 is described in Table \ref{tbl:ToyModelSetting}. At low temperatures (T$<$8.32 K), two vortices remain pinned. As the temperature increases beyond 8.32 K, one of the pinned vortices becomes depinned and escapes from the simulation due to the vortex-vortex repulsive force exceeding the pinning force provided by the pinning sites. This depinning event causes a noticeable downward jump in $P_\mathrm{2f}(T)$ around 8.32 K, as shown in Fig. \ref{fig:TDGLP2fJumpingDown}.


\section{Subtraction of probe background}
\label{sec:ProbeBackground}

Since the measured $P_\mathrm{2f}$ is the superposition of the probe background and the sample signal, the total measured signal may be weaker than the probe background if these two components are out of phase \cite{mircea2009phase}. In the absence of phase information, we perform a simple scalar subtraction of the probe background from the total signal. This naive background subtraction can result in negative $P_\mathrm{2f}$ values.


\section{Hysteresis in $P_\mathrm{2f}$ in temperature sweeps}
\label{sec:HysteresisCheck}


\begin{figure}
\includegraphics[width=0.45\textwidth]{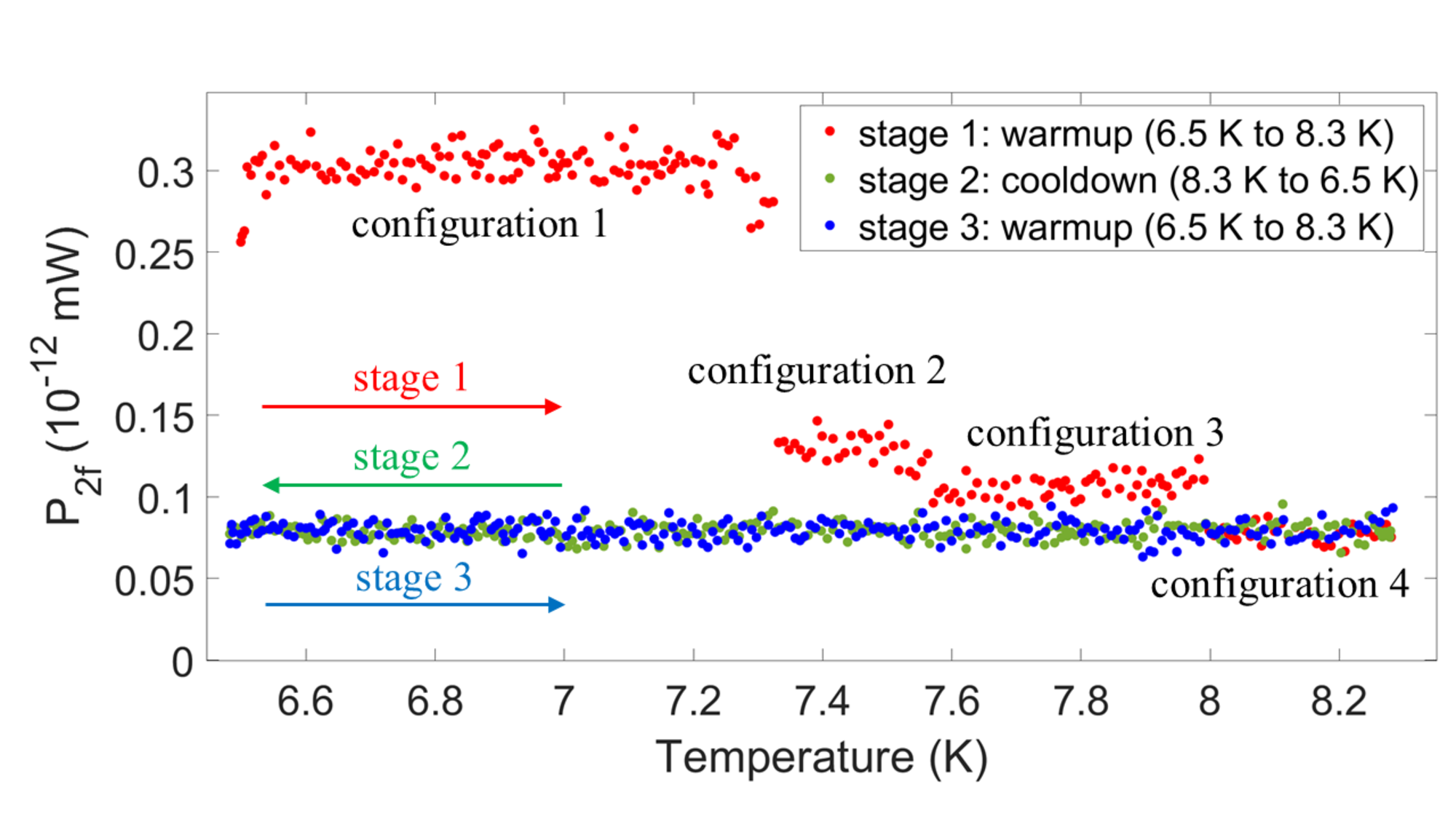}
\caption{\label{fig:CheckHysteresis1}A 3-stage zigzag temperature sweep measurement of $P_\mathrm{2f}$ at a fixed location on the sample surface. The input frequency is 1.818 GHz, and the input power is 2 dBm.}
\end{figure}

\begin{figure*}[t]
\includegraphics[width=\textwidth]{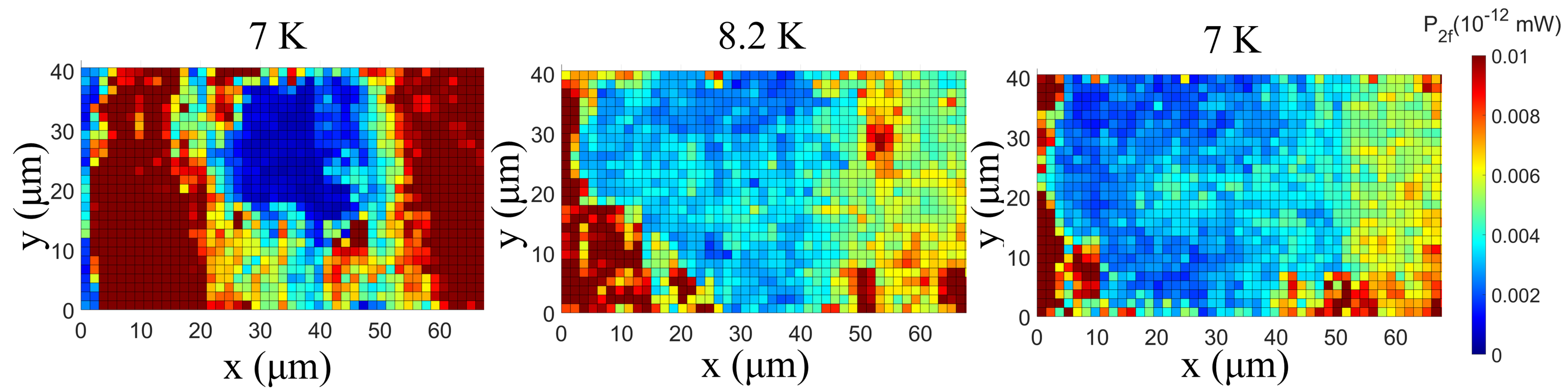}
\caption{\label{fig:Hysteresis2D}Three two-dimensional images of $P_\mathrm{2f}(x,y)$ measured on the Nb film, plotted with a common color-bar (right). The step size is $1.5 \, \mu$m. From left to right, the images are obtained successively at 7 K, 8.2 K, and then back to 7 K. The input frequency is 1.81 GHz, and the input power is 6.8 dBm.}
\end{figure*}

Figure \ref{fig:CheckHysteresis1} shows a 3-stage zigzag temperature sweep measurement of $P_\mathrm{2f}$. The sample is first cooled from 10 K to 6.5 K with $B_\mathrm{cooldown} = 1.7 $ mT. The DC magnetic field is then removed, and three successive microwave measurements are performed: stage 1 involves warming from 6.5 K to 8.3 K (red), stage 2 involves cooling back from 8.3 K to 6.5 K (green), and stage 3 involves a second warming from 6.5 K to 8.3 K (blue). Note that the sample remains below $T_{c}$ in this entire $P_\mathrm{2f}$ measurement process. In stage 1, three discrete jumps in $P_\mathrm{2f}(T)$ are observed, whereas $P_\mathrm{2f}(T)$ remains nearly temperature-independent during stages 2 and 3. This 3-stage zigzag temperature sweep clearly demonstrates the hysteresis of $P_\mathrm{2f}$ when $P_\mathrm{2f}(T)$ jumps are present.

One possible interpretation is as follows: At the beginning of stage 1, the vortex configuration is metastable. As the temperature increases, the trapped vortices rearrange themselves into progressively more stable configurations (from configuration 1 to configuration 4), which is reflected in the $P_\mathrm{2f}(T)$ jumps. Configuration 4 is the most stable among the four configurations, and thus the vortices tend to remain in this configuration during stages 2 and 3. This hysteresis behavior of $P_\mathrm{2f}(T)$ jumps aligns with the hypothesis that these jumps correspond to transitions of trapped vortices from less stable to more stable configurations.


As additional evidence of the hysteresis behavior of $P_\mathrm{2f}$, Fig. \ref{fig:Hysteresis2D} presents spatially-scanned images of $P_\mathrm{2f}$ at multiple temperatures. The sample is first cooled from 10 K to 7 K with $B_\mathrm{cooldown} = 1.7 $ mT. The DC magnetic field is then removed, and three successive microwave measurements are performed: the first image at 7 K (left), the second image after warming from 7 K to 8.2 K$<T_{c}$ (center), and the third image after cooling back from 8.2 K to 7 K (right). The contrast between the left and right images clearly demonstrates the hysteresis behavior of $P_\mathrm{2f}$.


\bibliography{P2fpaper.bib}

\end{document}